\documentclass[aps,pre,twocolumn,nofootinbibs,tightenlines,superscriptaddress]{revtex4-2}

\usepackage[utf8]{inputenc}
\usepackage[T1]{fontenc}

\usepackage{lmodern}
\usepackage{amsmath,amsfonts,amssymb}
\usepackage{mathtools}
\usepackage{graphicx}

\usepackage{lipsum}
\usepackage{comment}

\usepackage{hyperref}

\usepackage[usenames,dvipsnames]{xcolor}
\hypersetup{colorlinks=true, linkcolor=blue!50!black, urlcolor=blue!50!black, citecolor=blue!50!black}
\usepackage[all]{hypcap}

\usepackage{xcolor}

\usepackage{booktabs}

\usepackage{tikz}

\newcommand{\img}{\mathsf{i}}
\newcommand\diff{\mathrm{d}}

\newcommand{\subref}[2]{\hyperref[#1]{#2}}

\usepackage[normalem]{ulem}

\usetikzlibrary{matrix}
\usetikzlibrary{calc,fit}
\tikzset{%
  highlight1/.style={rectangle,rounded corners,color=red!,fill=red!15,draw,fill opacity=0.5,thick,inner sep=0pt}
}
\tikzset{%
  highlight2/.style={rectangle,rounded corners,color=green!,fill=green!15,draw,fill opacity=0.5,thick,inner sep=0pt}
}

\begin{document}

%\title{Time-dependent dynamics of the driven lattice Lorentz gas in three dimensions}
\title{Time-dependent dynamics in the confined lattice Lorentz gas}


%\author{Sebastian Leitmann}
%\thanks{}

\author{Alessio Squarcini}
\email[]{alessio.squarcini@uni.lu}
\affiliation{Institut f\"ur Theoretische Physik, Universit\"at Innsbruck, Technikerstra{\ss}e~21A, A-6020 Innsbruck, Austria}
%\email[]{alessio.squarcini@uibk.ac.at}

\affiliation{Complex Systems and Statistical Mechanics, Department of Physics and Materials Science, University of Luxembourg, 30 Avenue des Hauts-Fourneaux, L-4362 Esch-sur-Alzette, Luxembourg}

\author{Antonio Tinti}
\affiliation{Dipartimento di Ingegneria Meccanica e Aerospaziale, \\ Sapienza Universit\`a di Roma, via Eudossiana 18, 00184 Rome, Italy}
%\email[]{antonio.tinti@uniroma1.it}

\affiliation{Laboratory of Molecular Simulation (LSMO), Institut des Sciences et Ing{\'e}nierie Chimiques, {\'E}cole Polytechnique F{\'e}d{\'e}rale de Lausanne (EPFL), 1950, Sion, Switzerland}

\author{Pierre Illien}
\affiliation{Sorbonne Universit\'e, CNRS, Laboratoire PHENIX (Physico-Chimie des Electrolytes et Nanosyst\`emes Interfaciaux), 4 Place Jussieu, 75005 Paris, France}

\author{Olivier B\'enichou}
\affiliation{Laboratoire de Physique Th{\'e}orique de la Mati{\`e}re Condens{\'e}e, CNRS/UPMC, 4 Place Jussieu, F-75005 Paris, France}

\author{Thomas Franosch}
\email[]{thomas.franosch@uibk.ac.at}
\affiliation{Institut f\"ur Theoretische Physik, Universit\"at Innsbruck, Technikerstra{\ss}e~21A, A-6020 Innsbruck, Austria}

\date{\today}

\begin{abstract}
We study a lattice model describing the non-equilibrium dynamics emerging from the pulling of a tracer particle through a disordered medium occupied by randomly placed obstacles. The model is considered in a restricted geometry pertinent for the investigation of confinement-induced effects. We analytically derive  exact results for the characteristic function of the moments valid to first order in the obstacle density. By calculating the velocity autocorrelation function and its long-time tail we find that already in equilibrium the system exhibits a dimensional crossover. This picture is further confirmed by the approach of the drift velocity to its terminal value attained in the non-equilibrium stationary state. At large times the diffusion coefficient is affected by both the driving and confinement in a way that we quantify analytically. The force-induced diffusion coefficient depends sensitively on the presence of confinement. The latter is able to modify qualitatively the non-analytic behavior in the force observed for the unbounded model. We then examine the fluctuations of the tracer particle along the driving force. We show that in the intermediate regime superdiffusive anomalous behavior persists even in the presence of confinement. Stochastic simulations are employed in order to test the validity of the analytic results, exact to first order in the obstacle density and valid for arbitrary force and confinement.
\end{abstract}

%\pacs{}
%\keywords{}

\maketitle

%\section{}
%\subsection{}
%\subsubsection{}

%\tableofcontents

%%%%%%%%%%%%%%%%%%%%%%%%%%%%%%%%%%%%%%%%%%%%%%%%%%%%%%%%%%
%%%%%%%%%%%%%%%%%%%%%%%%%%%%%%%%%%%%%%%%%%%%%%%%%%%%%%%%%%
%\section{Introduction.}
%{\color{red}{\lipsum[1-3]}}

%%%%%%%%%%%%%%%%%%%%%%%%%%%%%%%%%%%%%%%%%%%%%%%%%%%%%%%%%%
%%%%%%%%%%%%%%%%%%%%%%%%%%%%%%%%%%%%%%%%%%%%%%%%%%%%%%%%%%

%%%%%%%%%%%%%%%%%%%%%%%%%%%%%%%%%%%%%%%%%%%%%%%%%%%%%%%%%%
%%%%%%%%%%%%%%%%%%%%%%%%%%%%%%%%%%%%%%%%%%%%%%%%%%%%%%%%%%

%%%%%%%%%%%%%%%%%%%%%%%%%%%%%%%%%%%%%%%%%%%%%%%%%%%%%%%%%%%%%%%%%%%%%
%%%%%%%%%%%%%%%%%%%%%%%%%%%%%%%%%%%%%%%%%%%%%%%%%%%%%%%%%%%%%%%%%%%%%
\section{Introduction}
Experiments on active-microrheology are nowadays largely employed in order to infer properties of soft materials \cite{Squires_2008, WP_2011, PV_2014}. In these experiments, a tracer particle is pulled through a medium by means of either optical or magnetic tweezers. The application of this experimental procedure finds applications in several contexts ranging from colloidal suspensions \cite{PV_2014}, soft glassy materials \cite{Bocquet_2008,Bocquet_2012}, fluid interfaces \cite{Choi:2011gz}, to living cells \cite{HLRG_1999}. The possibility to tune the external force allows for the exploration of the full crossover from the linear regime to the truly non-linear one emerging for strong forces and its associated non-equilibrium phenomena such as force-thinning \cite{HSLW_2004, CB_2005, SMF_2010} and enhanced diffusivities \cite{WHVB_2012, WH_2013, Senbil_2019}. In addition to these phenomena, among the most fascinating challenges for the theory is the description of effects due to biasing, crowding, and confinement \cite{Hanggi, BIOSV_2018} in strongly interacting systems out of equilibrium.

Several analytical results have been obtained for lattice models in which a tracer particle moves in a dense environment \cite{Benichou_2013, Benichou_2014, Benichou_2016}. In particular, for single-file diffusion there exist exact \cite{Illien_2013, Illien_2014} and numerical results \cite{Jack_2008, Basu_2014, BSV_2015, SN_2018} as well as approximate \cite{Illien_2015} and experimental ones \cite{Siems_2012}, including also the theoretical analysis of the biased dynamics \cite{MB_1991} and the motion in an external potential \cite{BS_2009}. Furthermore, analytical progresses have been made also on closely related fronts of the literature. This includes -- inter alia -- exclusion processes \cite{GM_2006, Mallick_2011, KMS_2014, IMS_2017} and their variants with a moving defect \cite{Mallick_1996, LEM_2022}, driven one-dimensional lattice gases \cite{HMS_1999, KLMT_2022} in narrow channels \cite{CMP_2017, MMP_2020, MM_2021, CMP_2018, MMP_2021}, diffusion on percolation networks \cite{Barma_Dhar_1983, Dhar_1984, Bowditch_Croydon_2022}\footnote{See \cite{Fribergh_Hammond_2013} for a rigorous proof of the results of \cite{Barma_Dhar_1983}.}, and the random trapping for a tracer moving into a channel \cite{Shafir_2024}. Despite these achievements, an analytically tractable model able to encompass the out-of-equilibrium regime of strong pulling in the presence of crowing, disorder, and geometrical confinement is still missing to date.

This paper aims at providing first attempt in this direction through the driven quasi-confined lattice Lorentz gas model. This model describes the complementary situation in which a tracer particle explores an immobile landscape of randomly distributed obstacles. This minimal model of disorder is motivated by the fact that the mutual exclusion between the tracer and the environment is commonly regarded as the most important ingredient. The confinement in a lattice model can be realized in several ways: the simplest is referred to as \emph{quasi-confinement}, which is implemented by restricting the model onto an infinitely long strip with identified edges; this choice corresponds to the topology of a cylinder of finite circumference $L$. 

For this lattice model we shall elaborate an analytical solution, valid to first order in the obstacle density, $n$, and for arbitrarily strong driving and confinement size. In this regard, all formulas should be read including order $O(n)$. Strictly speaking, this approach cannot describe finite $n$, a regime that is far beyond the scope of the paper; hence, conclusions cannot be drawn for finite densities at infinite times. To this end, it is important to stress a further intricacy.  The limits of $n \rightarrow 0$ and $F \rightarrow \infty$ (with $t \rightarrow \infty$) \emph{do not} commute, this makes the scenario rather subtle. However, the simulations we performed show in a clear fashion the domain of applicability of the theory and, anticipating some results, we specify that the window of validity is \emph{not} uniform as it depends on $F$ and $n$. We therefore compute the first-order correction for arbitrary force and then, for practical purposes, we compare it with simulations protracting to large although finite times, a circumstance that applies to practical scenarios. The theory encompasses the full crossover from the maximally confined $L=2$ lane model to the unbounded model ($L=\infty$) whose solution in equilibrium is known for a long time by the pioneering works of Nieuwenhuizen \emph{et al.}\@ in Refs.~\cite{OK_nieuwenhuizen_3, OK_nieuwenhuizen_2, OK_nieuwenhuizen_4, OK_nieuwenhuizen_1, VANLEEUWEN1967457, WEIJLAND196835, ernst1971long}, while the non-equilibrium case has been solved more recently \cite{LF_2013, LF_2017, LSF_2018, LSF_2018, LBF_2018, Squarcini_2024_LLG_discrete_time}. The analytical method, developed in \cite{LF_2013, LF_2017, LSF_2018}, and here applied to the quasi-confined model, relies on a lattice version of the scattering formalism borrowed from quantum mechanics \cite{Ballentine}.

The scope of this paper is to derive and discuss the complete time-dependent dynamics, the approach to the non-equilibrium stationary state, and how these aspects interplay with the presence of a confinement. The principle results of this paper can be summarized as follows: the stationary state is characterized by a terminal velocity which is approached -- for infinitesimally small forces -- as the power-law $t^{-1/2}$ for \emph{any finite} value of $L$, whereas the power law is $t^{-1}$ for the unbounded model \cite{LF_2013}. The above long-time tails are actually \emph{fragile} since an arbitrarily small but finite force is sufficient to decorate them by an exponential decay; i.e., for the confined system the long time-tail becomes $t^{-1/2} \exp(-c F^{2}t)$, where $c>0$ is a coefficient. The existence of an effective dimensionality is actually the signature of a \emph{dimensional crossover}. This surprising behavior, that we find already at equilibrium, can be characterized by the complete time-dependent velocity autocorrelation function (VACF). The latter exhibits a long-time tail of the form $t^{-(d+2)/2}$ \cite{ernst1971long} with effective dimension $d=2$ at short times, while for large times the effective dimension $d=1$ entails a reduction to a quasi one-dimensional system. We then study the fluctuations of the tracer particle along the force. From the variance of the position we infer the emergence of a transient superdiffusive behavior at intermediate times with a time-dependent exponent of anomalous diffusion that can saturates to $\alpha(t)=3$ for large forces. Standard diffusion, i.e., $\alpha(t)=1$, is attained at both short times and long times. At long times however the diffusion is characterized by a force-dependent and confinement-dependent diffusion coefficient that we identify. Among our findings is that, in general, for large forces the confinement effects fade out and the system behaves, \emph{de facto}, as the unbounded one. On the other hand, confinement-induced effects play a major role at low forces. Their effect is so strong that it can alter the non-analytic behavior of response functions making them significantly different from the one observed in the unbounded model. This is what we observed by studying the long-time diffusion coefficient.

This paper is organized as follows. In Sec.~\ref{sec_model} we introduce the driven, quasi-confined, lattice Lorentz gas model. The exact solution of the model relies on the scattering approach formalism elaborated in \cite{LF_2013, LF_2017, LSF_2018} that we briefly recall in Sec.~\ref{sec_solution}. The results for the velocity response are presented in Sec.~\ref{sec_velocity_relaxation}. Section~\ref{sec_fluctuations} conveys the analysis of the VACF in equilibrium and its associated dimensional crossover, as well as the time-dependent fluctuations along the force. Our summary and conclusions are provided in Sec.~\ref{sec_summary_and_conclusions}. A series of Appendices collects several technicalities including Green's function of the quasi-confined model.

%%%%%%%%%%%%%%%%%%%%%%%%%%%%%%%%%%%%%%%%%%%%%%%%%%%%%%%%%%%%%%%
%%%%%%%%%%%%%%%%%%%%%%%%%%%%%%%%%%%%%%%%%%%%%%%%%%%%%%%%%%%%%%%
\section{The model}
\label{sec_model}
\begin{figure}[htbp]
\centering
\includegraphics[width=8.2cm]{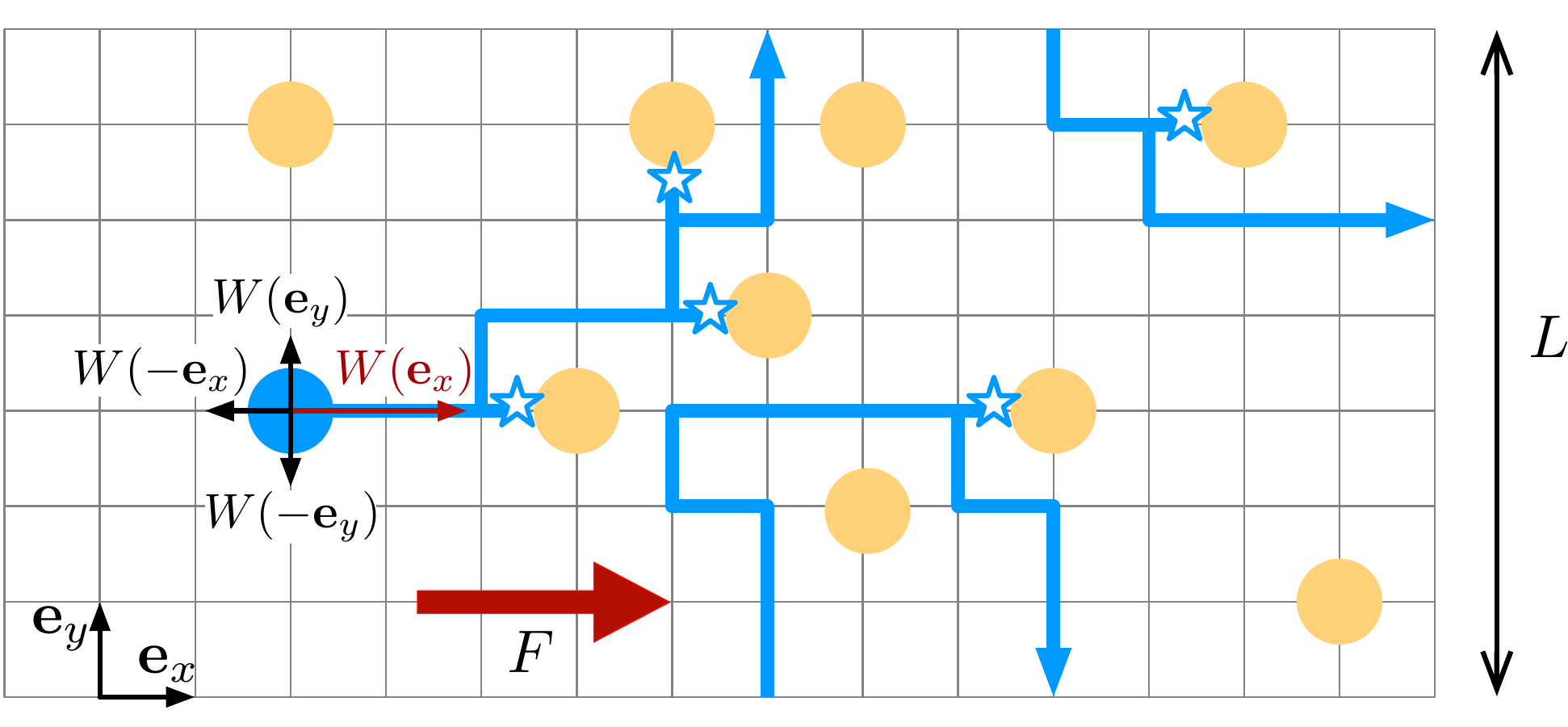}
\caption{The tracer particle performs a random walk under the action of a force $F$ resulting in biased hopping rates (or nearest neighbor transition probabilities). Rejected transitions onto obstacles (bullets in the picture) are indicated with a star.}
\label{Fig_1_cylinder}
\end{figure}

The lattice Lorentz gas model is a random walk occurring on a disordered lattice $\Lambda$ in which some sites are not accessible to the walker. Let $\Lambda = \{ \mathbf{r} = (x,y): x = 1,\ldots, L_{x} , y=1,\ldots,L_y \}$ with $N = L_{x} L_{y}$ sites and periodic boundary conditions. The walker performs jumps to its nearest neighbors, $\mathcal{N} = \{ \pm \textbf{e}_{x}, \pm \textbf{e}_{y} \}$. The waiting time at accessible sites is exponentially distributed with mean waiting time $\tau$. The disorder is modeled via hard and immobile obstacles (also referred to as impurities). If the tracer attempts to hop on a site occupied by an obstacle, then the transition is rejected and the walker remains where it was (still time is progressing and the walker doesn't move); the trajectory is schematically depicted in Fig.~\ref{Fig_1_cylinder}. Throughout this paper we will consider periodic boundary conditions along both the lattice directions. Technically speaking, the limit $L_{x} \rightarrow \infty$ is anticipated while the size $L\coloneq L_y $ will be kept finite. This procedure defines the notion of quasi-confinement adopted in this paper.

The driving is implemented via a force $F \geqslant 0$, i.e., we consider only positive forces that pull the tracer along the $x$-axis, as sketched in Fig.~\ref{Fig_1_cylinder}. For the sake of convenience, the force $F$ is dimensionless in our calculation, it is measured in units of $k_{B}T/a$, where $a$ is the lattice spacing that we set equal to unity. The bias introduced by the force affects the transition probabilities for hopping on neighboring sites. Imposing local detailed balance implies that $W(\textbf{e}_{x})/W(-\textbf{e}_{x})=\exp(F)$ and $W(\textbf{e}_{y})/W(-\textbf{e}_{y})=1$ \cite{HK_1987}, therefore
\begin{eqnarray}
W(\pm\textbf{e}_{x}) & = & \frac{e^{\pm F/2}}{2+e^{F/2}+e^{-F/2}} \\
W(\pm\textbf{e}_{y}) & = & \frac{1}{2+e^{F/2}+e^{-F/2}} \, .
\end{eqnarray}
These expression follow by the above criterion as well as the overall normalization. As a result, for large $F  \gg 1$ the dominant transition rate is $W(\textbf{e}_{x})$, the latter is enhanced at the expenses of the transitions in the direction perpendicular and opposite to the force. Note that in the absence of bias ($F=0$) all transition probabilities become identical and equal to $1/4$.

The state of the system at time $t$ is completely characterized by the site-occupation probability density $p_{\textbf{r}}(t)$ which expresses the probability to be at site $\textbf{r}$ at time $t$. We formally introduce a Hilbert space $\mathcal{H}$ with orthonormal basis states of position kets $\{ \vert \textbf{r} \rangle: \textbf{r} \in \Lambda \}$. We then define an abstract ket state $| p(t) \rangle \in \mathcal{H}$ 
by $\vert p(t) \rangle = \sum_{ \textbf{r} \in \Lambda} p_\textbf{r}(t) \vert \textbf{r} \rangle$. Using completeness we find
\begin{equation}
\label{ }
\vert p(t) \rangle = \sum_{ \textbf{r} \in \Lambda} \vert \textbf{r} \rangle \langle \textbf{r} \vert p(t) \rangle \, .
\end{equation}
Hence, $p_{\textbf{r}}(t) = \langle \textbf{r} | p(t) \rangle$ is the probability to find the tracer at site $\textbf{r}$ at time $t$. The time evolution of this probability follows according to the master equation $\partial_{t} \vert p(t) \rangle = \hat{H} \vert p(t) \rangle$. The latter is formally analogous to the Schr\"odinger equation in imaginary time provided $\hat{H}$ is identified as the ``Hamiltonian''. Upon projecting the master equation onto the bra $\langle \textbf{r} \vert$ we find the following representation of the master equation in the position basis
\begin{equation}
\label{12122023_1533}
\partial_{t} \langle \textbf{r} \vert p(t) \rangle = \sum_{ \textbf{r} \in \Lambda } \langle \textbf{r} \vert \hat{H} \vert \textbf{r}^{\prime} \rangle \langle \textbf{r}^{\prime}  \vert p(t) \rangle \, ,
\end{equation}
from which we can identify the matrix elements $\langle \textbf{r} \vert \hat{H} \vert \textbf{r}^{\prime} \rangle$ as the transition rates from site $\textbf{r}^{\prime}$ to $\textbf{r}$.

The motion on the empty lattice, i.e., the lattice with no obstacles, can be regarded as the reference situation corresponding to the unperturbed Hamiltonian
\begin{equation}
\label{12122023_1534}
\hat{H}_{0} = \frac{\Gamma}{\tau} \sum_{ \textbf{r} \in \Lambda } \biggl[ - \vert \textbf{r} \rangle \langle \textbf{r} \vert + \sum_{ \textbf{d} \in \mathcal{N} } W(\textbf{d}) \vert \textbf{r} \rangle \langle \textbf{r}-\textbf{d} \vert \biggr] \, ,
\end{equation}
with non-normalized rates $(\Gamma/\tau) W(\textbf{d} \in \mathcal{N})$, where
\begin{equation}
\label{gamma_rate}
\Gamma = \frac{1+\cosh(F/2)}{2} \, .
\end{equation}
The choice of this prefactor corresponds to non-normalized transition rates considered in \cite{LF_2017, LSF_2018}, while normalizad rates have been considered in \cite{LF_2013}. The two choices are actually equivalent under a rescaling of the time. One may choose units of time such that the mean waiting time $\tau=1$ sets the time scale. The presence of hard immobile obstacles precluding transitions \emph{from} and \emph{to} impurities can be modeled by adding an interaction potential that cancels exactly the associated transition rates. The Hamiltonian is formally written as $\hat{H} = \hat{H}_{0} + \hat{V}$, where the ``potential'' $\hat{V}$ is constructed such that its matrix elements cancel the non-allowed transitions. For a single obstacle located at site $\textbf{s}$ the potential takes the form $\hat{V}=\hat{v}(\textbf{s}) \equiv \hat{v}_{1}$. The nonvanishing matrix elements are those between the state associated to the site $\textbf{s}$ and its neighbors: $\textbf{s} - \textbf{d}$, with $\textbf{d} \in \mathcal{N}$. For example, $\langle \textbf{s} \vert \hat{v}_{1} \vert \textbf{s}-\textbf{d} \rangle = -\Gamma W(\textbf{d})$ cancels the transition \emph{to} the impurity, while $\langle \textbf{s}-\textbf{d} \vert \hat{v}_{1} \vert \textbf{s}\rangle = -\Gamma W(-\textbf{d})$ cancels the transition \emph{from} the impurity, and $\langle \textbf{s}-\textbf{d} | \hat{v}_1 | \textbf{s}-\textbf{d} \rangle = \Gamma W(\textbf{d})$.

%%%%%%%%%%%%%%%%%%%%%%%%%%%%%%%%%%%%%%%%%%%%%%%%%%%%%%%%%%%%%%%%%%%%%
%%%%%%%%%%%%%%%%%%%%%%%%%%%%%%%%%%%%%%%%%%%%%%%%%%%%%%%%%%%%%%%%%%%%%
\section{Solution strategy}
\label{sec_solution}
In this Section we outline the main ideas underlying the analytical approach for the solution in the presence of disorder. To begin, we illustrate how to solve the reference case corresponding to the driven dynamics in the absence of obstacles. The material presented in this Section transforms the established solution strategy for the unbounded model \cite{LF_2013, LF_2017, LSF_2018} to the quasi-confined model.

%%%%%%%%%%%%%%%%%%%%%%%%%%%%%%%%%%%%%%%%%%%%%%%%%%%%%%%%%%%%%%%%%%%%%%
%%%%%%%%%%%%%%%%%%%%%%%%%%%%%%%%%%%%%%%%%%%%%%%%%%%%%%%%%%%%%%%%%%%%%%
\subsection{Bare dynamics}
The random walk is described in terms of the conditional probability $U_0(\textbf{r}, t|\textbf{r}')$ of the tracer to move from site $\textbf{r}'$ to site $\textbf{r}$ in lag time $t$. The time evolution of the conditional probability $U_0(\textbf{r},t |\textbf{r}')$ is governed by a master equation of the form
\begin{eqnarray} \nonumber
\partial_t U_0(\textbf{r},t|\textbf{r}') & = & \sum_{\textbf{d}\in \mathcal{N}} W(\textbf{d}) U_0(\textbf{r}-\textbf{d}, t | \textbf{r}') \\
& - & \sum_{\textbf{d}\in \mathcal{N}} W(\textbf{d}) U_0(\textbf{r}, t | \textbf{r}') \, .
\end{eqnarray}
By adopting the bra-ket notation the conditional probability is written as $U_0(\textbf{r}, t|\textbf{r}') = \langle \textbf{r} | \hat{U}_0(t) |\textbf{r}' \rangle$ in the complete orthonormal basis of position kets with time-evolution operator $\hat{U}_0(t)$. The master equation for the time-evolution operator reads $\partial_t \hat{U}_0(t) = \hat{H}_0 \hat{U}_0(t)$, with initial condition $\hat{U}_0(t=0) = \openone$ and $\hat{H}_{0}$ given by Eq.~(\ref{12122023_1534}). The time-evolution operator formally solves the master equation by means of $\hat{U}_0(t) = \exp(\hat{H}_0 t)$. The site-occupation probability at time $t$ given that the driving force $F>0$ is switched on at time $t=0$ is given by
\begin{equation}
\label{ }
\vert p(t) \rangle = \hat{U}_{0}(t) \vert p_{\rm eq} \rangle \, ,
\end{equation}
where $\vert p(t=0) \rangle = \vert p_{\rm eq} \rangle = (1/N)\sum_{\textbf{r}\in\Lambda} |\textbf{r}\rangle$ represents the probability distribution at time $t=0$ consisting of equally likely lattice sites. The time evolution operator can be explicitly calculated by considering the plane-wave basis
\begin{align} \label{eq:definition_plane_wave_basis}
| \textbf{k} \rangle =\frac{1}{\sqrt{N}} \sum_{\textbf{r}\in\Lambda }  e^{\img \textbf{k}\cdot \textbf{r}} |\textbf{r} \rangle \, ,
\end{align}
with wave vector $\textbf{k}$ defined such that $\textbf{k} = (k_x, k_y) \in \Lambda^* := \{(2\pi p/L_x, 2\pi q/L) : p = 0,\ldots, L_x, q = 0, \ldots, L\}$. Since the unperturbed Hamiltonian $\hat{H}_0$ is translationally invariant, it is diagonal in the plane-wave basis~[Eq.~\eqref{eq:definition_plane_wave_basis}] and therefore
\begin{align}
\langle \textbf{k}|\hat{H}_0|\textbf{k}^{\prime}\rangle = \epsilon(\textbf{k}) \delta_{\textbf{k},\textbf{k}^{\prime}} \, ,
\end{align}
where $\delta_{\textbf{k},\textbf{k}^{\prime}}$ denotes the Kronecker delta and the eigenvalue $\epsilon(\textbf{k})$ is given by
\begin{equation}
\label{ }
\epsilon(\textbf{k}) = \Gamma \biggl[-1 + \sum_{\textbf{d}\in\mathcal{N}} W(\textbf{d}) e^{-\img\textbf{k}\cdot\textbf{d}}\biggr] \, ,
\end{equation}
or equivalently,
\begin{equation}
\label{ }
\epsilon(\textbf{k}) = - \Gamma \sum_{\textbf{d}\in\mathcal{N}} \Bigl[ 1 - \cos(\textbf{k}\cdot\textbf{d}) + \img \sin(\textbf{k}\cdot\textbf{d}) \Bigr] W(\textbf{d}) \, .
\end{equation}
In turn, this can be cast in the following way
\begin{equation}
\label{ }
\epsilon(\textbf{k}) = \epsilon_{x}(k_{x}) + \epsilon_{y}(k_{y}) \, ,
\end{equation}
where
\begin{equation}
\begin{aligned}
\label{}
\epsilon_{x}(k) & = - \frac{1}{2} \biggl[ (1-\cos(k))\cosh(F/2) +  \img \sin(k)\sinh(F/2) \biggr] \\
\epsilon_{y}(k) & = \epsilon_{x}(k)\big\vert_{F=0} \, .
\end{aligned}
\end{equation}
Quite interestingly, the effect of a nonzero force amounts to an imaginary shift of the momentum from $k_{x}$ to $k_{x}+\img F/2$ and thus $\epsilon_{x}(k)$ in the above equation can be written in the equivalent form
\begin{equation}
\label{ }
\epsilon_{x}(k) = \frac{1}{2} [1-\cosh(F/2)] - \frac{1}{2}[1-\cos(k+\img F/2)] \, .
\end{equation}

The characteristic function
\begin{equation}
\label{ }
F_0(\textbf{k},t) = \langle e^{-\img\textbf{k}\cdot\Delta\textbf{r}(t)}\rangle_0
\end{equation}
encodes all moments of the displacement $\Delta \textbf{r}(t) = \textbf{r}(t) - \textbf{r}(0) = (\Delta x(t), \Delta y(t))$ in the absence of obstacles, as emphasized with the notation $\langle \cdot\rangle_0$. In terms of the time-evolution operator, the above can be evaluated to
\begin{equation}
\begin{aligned}
F_0(\textbf{k}, t) & = \sum_{\textbf{r},\textbf{r}'\in\Lambda} e^{-\img\textbf{k}\cdot(\textbf{r}-\textbf{r}')}\langle \textbf{r}|\hat{U}_0(t)|\textbf{r}'\rangle \langle\textbf{r}'|p_\text{eq}\rangle \, ,
\end{aligned}
\end{equation}
where $\langle \textbf{r}^{\prime} \vert p_{\text{eq}} \rangle = 1/N$ is the initial distribution. The eigenvalue of the time evolution in the plane wave basis determines the intermediate scattering function via
\begin{equation}
\begin{aligned}
F_0(\textbf{k}, t) & =  \langle \textbf{k}| \hat{U}_0(t)|\textbf{k}\rangle = \exp[\epsilon(\textbf{k})t] \, .
\end{aligned}
\end{equation}
The first moment is 
\begin{align}
\label{05032024_1508}
\langle \Delta x(t)\rangle_0 = \img \frac{\partial}{\partial k_x} F_0(\textbf{k},t)\Bigr|_{\textbf{k}=0}  = v_{0} t \, ,
\end{align}
with drift velocity $v_{0} = \img \partial \epsilon(\textbf{k})/\partial k_{x} \vert_{ \textbf{k}=0 } = \sinh(F/2)/2$. For the second moment along the $x$ direction, we calculate
\begin{align}
\label{18092023_1424}
\langle \Delta x^2(t)\rangle_0 = -\frac{\partial^2}{\partial k_x^2} F_0(\textbf{k},t)\Bigr|_{\textbf{k}=0}  = (v_{0}t)^{2} +  2 D_{x}^{0} t \, ,
\end{align}
where $D_{x}^{0}$ is the bare diffusion coefficient
\begin{equation}
\label{09122025_1040}
D_{x}^{0}(F) = \frac{\cosh(F/2)}{4} \, .
\end{equation}
We note that as long as $L_{x}$ is finite the momenta $k_{x}$ form a discrete set but since the limit $L_{x}\rightarrow \infty$ is anticipated it follows that such momenta form a continuum; this allows us to perform the momentum derivatives $\partial_{k_{x}}$ as shown in Eq.~(\ref{18092023_1424}). Parenthetically, we observe that $D_{x}^{0}(F)$ does not depend on the value of the size $L$.

%%%%%%%%%%%%%%%%%%%%%%%%%%%%%%%%%%%%%%%%%%%%%%%%%%%%%%%%%%%%%%%%%%%%%
%%%%%%%%%%%%%%%%%%%%%%%%%%%%%%%%%%%%%%%%%%%%%%%%%%%%%%%%%%%%%%%%%%%%%
\subsection{Obstructed biased motion}

We consider the presence of hard, immobile obstacles randomly distributed on the lattice. The effects due the disordered lattice can be worked out exactly to first order in the obstacle density $n$ and for an arbitrary strength $F$ of the bias. In this section we  review the analytical technique based on the application of the scattering formalism in quantum mechanics (see e.g. \cite{Ballentine}) and originally developed in \cite{LF_2013, LSF_2018, LSF_2018} for the lattice Lorentz gas.

When the tracer attempts to jump onto an obstacle site, the tracer remains at its initial site before the jump. Thus, the obstacle cancels the transition from and to the unaccessible site. In order to take into account the obstacles, we split the Hamiltonian into the sum of a free contribution $\hat{H}_{0}$ describing the dynamics on the empty lattice and an interaction term $\hat{V}$ accounting for the disorder, i.e., $\hat{H}=\hat{H}_{0}+\hat{V}$. Let $\hat{v}(\textbf{s}_i) \equiv \hat{v}_i$ be the potential for a single impurity located at site $\textbf{s}_i$.  Let $N_I$ be the number of such obstacles on the lattice, then the total potential $\hat{V}$ can be written as a sum over all single-obstacle potentials $\hat{v}_i$ with $\hat{V} = \sum_{i = 1}^{N_I} \hat{v}_i$, the density of the obstacle is thus $n= N_I/N$. Technically, a tracer beginning on an empty site will remain within the connected cluster of empty sites.

Let us consider now an isolated obstacle that we put in site $\textbf{s}_{1}$. However, since the system is translationally invariant, we may pick any site as origin, so we do for $\textbf{s}_{1}$. Since the tracer performs nearest-neighbor jumps, the only non-vanishing matrix elements for an obstacle located in the origin are the obstacle site $\textbf{e}_0 = \textbf{0}$ and its nearest neighbors $ \textbf{e}_{\pm 2} = \pm \textbf{e}_y$, $\textbf{e}_{\pm1}= \pm \textbf{e}_x$. Cancellation of the transition rates entails that matrix elements $\langle \textbf{e}_\mu |\hat{v}|\textbf{e}_\nu\rangle$ of the single-obstacle potential are $\langle \textbf{0} \vert \hat{v} \vert \textbf{0}-\textbf{d} \rangle = - \Gamma W(\textbf{d})$ and $\langle \textbf{0}-\textbf{d} \vert \hat{v} \vert \textbf{0} \rangle = - \Gamma W(-\textbf{d})$ where $\textbf{d} \in \mathcal{N}$, where $\mathcal{N}$ denotes the neighboring sites of the obstacle (see the discussion at the end of Sec.\ref{sec_solution}). The matrix representation of the single obstacle potential in the basis $(\textbf{e}_{-2}, \textbf{e}_{-1}, \textbf{e}_0, \textbf{e}_1, \textbf{e}_2 )$ reads
\begin{align} \label{eq:matrix_obstacle_potential}
v = \frac{1}{4\tau} 
\begin{pmatrix}
  1 & 0 & -1 & 0 & 0  \\
  0 & e^{F/2} & - e^{-F/2} & 0 & 0  \\
  -1 & -e^{F/2} & 4\Gamma & -e^{-F/2} & -1 \\
  0 & 0 & -e^{F/2} & e^{-F/2} & 0  \\ 
  0 & 0 & -1 & 0 & 1  \\
\end{pmatrix} \, ,
\end{align}
where $\tau$ will be set equal to unity in all calculations. Probability conservation entails that all entries in each column sum up to zero.

With this representation, we can follow the approach outlined in Ref.~\cite{LSF_2018}. We express the dynamics of the tracer particle in the presence of obstacles in terms of the full propagator $\hat{G}$, the latter being defined as the Laplace transform of the time-evolution operator $\hat{U}(t)$ for a configuration of impurities with interaction potential $\hat{V}=\sum_{i=1}^{N_{I}} \hat{v}_{i}$, therefore
\begin{align}
\hat{G}(s) = (s-\hat{H})^{-1} \, .
\end{align}
The full propagator $\hat{G}(s) = (s-\hat{H})^{-1}$ and the free propagator $\hat{G}_{0}(s) = (s-\hat{H}_{0})^{-1}$, written as resolvents, are related via
\begin{equation}
\label{ }
\hat{G} = \hat{G}_{0}+\hat{G}_{0}\hat{V}\hat{G} \, ,
\end{equation}
where $\hat{G}_0$ is the bare propagator corresponding to the empty lattice. This equation admits an iterative solution in terms of the Born series, $\hat{G} = \hat{G}_{0} + \hat{G}_{0} \hat{V} \hat{G}_{0} + \hat{G}_{0} \hat{V} \hat{G}_{0} \hat{V} \hat{G}_{0} + \dots$. By employing the formalism of quantum mechanical scattering theory the propagator can be equivalently expressed as
\begin{align} \label{eq:propagator_via_scattering_operator}
\hat{G} = \hat{G}_0 + \hat{G}_0 \hat{T} \hat{G}_0 \, ,
\end{align}
in which the scattering operator $\hat{T}$ is related to the interaction potential via $\hat{T} = \hat{V}+\hat{V}\hat{G}_{0}\hat{T}$, the latter encodes all possible sequences of collisions of the tracer with the disorder. Analytic progress is made by rearranging the terms in the form of a multiple scattering expansion
\begin{align} \label{eq:scattering_expansion_t}
\hat{T} = \sum_{i=1}^{N_I} \hat{t}_i + \sum_{\substack{j,k=1\\ j\neq k}}^{N_I}\hat{t}_j \hat{G}_0 \hat{t}_k
+ \sum_{\substack{l,m,n=1 \\ l\neq m, m\neq n}}^{N_I} \hat{t}_l\hat{G}_0\hat{t}_m\hat{G}_0\hat{t}_n + \dotsb \, .
\end{align}
The first sum in Eq.~(\ref{eq:scattering_expansion_t}) takes into account the revisit of a certain impurity, this term contributes in first order of the density $n$. The remaining sums appearing in the multiple scattering expansion [Eq.~(\ref{eq:scattering_expansion_t})] correspond to contributions involving correlated scattering with at least two impurities, these terms contribute to order $O(n^{2})$ and higher. Throughout this paper we will not be interested in the effects due to a specific arrangement of the lattice impurities, therefore we will perform an average over the disorder realizations, this is denoted $[ . ]_{\rm av}$. The averaging procedure restores the translational invariance and, as a result, the disorder-averaged scattering operator $ [ \hat{T} ]_{\rm av}$ is diagonal in the plane-wave basis $\vert \textbf{k}\rangle$. The first term in the scattering expansion [Eq.~(\ref{eq:scattering_expansion_t})] is identified as the one yielding a linear contribution in the density $n$. The remaining terms are responsible for corrections of order $O(n^{2})$ or higher and are interpreted as effects arising from correlated scatterings between different impurities. Retaining the first sum in the scattering expansion amounts to the so-called single-scatterer approximation \cite{vRN_1999}. This approximation scheme yields $\langle \textbf{k} \vert [ \hat{T} ]_{\rm av} \vert \textbf{k} \rangle = nN \langle \textbf{k} \vert \hat{t} \vert \textbf{k} \rangle + O(n^{2})$, where $\langle \textbf{k} \vert \hat{t} \vert \textbf{k} \rangle := t(\textbf{k})$ is the forward scattering amplitude of a single obstacle and $nN = N_I$ is the number of impurities. From Eq.~(\ref{eq:propagator_via_scattering_operator}) we obtain the disorder-average propagator in plane wave basis, $[G]_\text{av}(\textbf{k}) := \langle \textbf{k} | [\hat{G}]_\text{av}| \textbf{k} \rangle$, exact to first order in the obstacle density
\begin{align} \label{eq:disorder_averaged_prop}
[G]_\text{av}(\textbf{k}) =  G_0(\textbf{k}) + n N t(\textbf{k}) G_0(\textbf{k})^2 + O(n^2) \, .
\end{align}
The first term in the equation above is the bare propagator in Fourier space and Laplace domain, therefore
$G_{0}( \textbf{k} ) = \langle \textbf{k} \vert \hat{G}_{0} \vert \textbf{k} \rangle = \int_{0}^{\infty}\textrm{d}t \, e^{-st} \langle \textbf{k} \vert \hat{U}_{0}(t) \vert \textbf{k} \rangle = [ s - \epsilon(\textbf{k})]^{-1}$, the latter is diagonal in the plane-wave basis. From now on the dependence on the Laplace parameter $s$ will be suppressed for notational convenience. The second term in Eq.~(\ref{eq:disorder_averaged_prop}) is the self-energy $\Sigma(\textbf{k}) = n N t(\textbf{k}) + O(n^{2})$ to linear order in the density $n$, the latter follows from Dyson's equation, $[G_{\rm av}](\textbf{k}) = \left[ G_{0}(\textbf{k})^{-1} - \Sigma(\textbf{k})\right]^{-1}$. 

\begin{figure}[htbp]
\centering
\includegraphics[width=0.5\columnwidth]{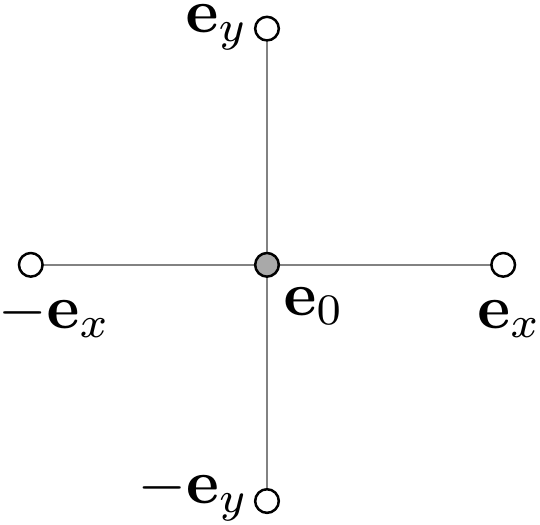}
\caption{The distinguished basis $\mathcal{B}$ centered on the impurity.}
\label{fig_basis}
\end{figure}

Within the theoretic formulation outlined above the tracer can start its motion at any lattice site, however in simulations the initial position is chosen at an accessible site. This prescription can be properly taken into account by correcting the disorder-averaged propagator by a factor $1/(1-n) = 1+n+O(n^2)$ and then retaining terms linear in $n$, therefore Eq.~(\ref{eq:disorder_averaged_prop}) is corrected by
\begin{equation}
\label{21092023_1631}
[G]_\text{av}^{\text c}(\textbf{k}) = G_{0}(\textbf{k}) + n [G_{0}(\textbf{k})+  N t(\textbf{k}) G_{0}(\textbf{k})^{2}] + O(n^{2}) \, .
\end{equation}
The disorder-averaged propagator encodes all moments of the tracer's displacement to first order in the obstacle density $n$ and is valid for arbitrarily strong driving. At linear order in $n$, one must include sequences in which the particle undergoes arbitrarily many scatterings at a given obstacle, then moves to a different obstacle and scatters there arbitrarily many times, and so on -- subject to the crucial constraint that no previously visited obstacle is revisited after the tracer has interacted with a different obstacle. Processes that violate this non-revisiting condition (i.e., returns to an earlier obstacle after encountering another) contribute only beyond first order in $n$ and are therefore excluded at this level. Consequently, the linear-in-$n$ scattering series is built from free propagations interspersed with single-obstacle scattering episodes, each of which is completely characterized by the single-impurity $t$-matrix. The single-obstacle scattering operator $\hat{t}$ satisfies
\begin{align}
\label{10112023_1057}
\hat{t} = \hat{v} + \hat{v} \hat{G}_0 \hat{t} = \hat{v} + \hat{t} \hat{G}_0 \hat{v} \, .
\end{align}
The only non-vanishing matrix elements for an obstacle located in the origin are the obstacle site $\textbf{e}_0 = \textbf{0}$ and its nearest neighbors $ \textbf{e}_{\pm 2} = \pm \textbf{e}_y, \textbf{e}_{\pm1}= \pm \textbf{e}_x$. To be definite, the distinguished basis is $\mathcal{B} = ( - \textbf{e}_{y}, - \textbf{e}_{x}, \textbf{e}_{0}, \textbf{e}_{x}, \textbf{e}_{y} )$; see Fig.~\ref{fig_basis}.
Since the potential $\hat{v}$ displays non-vanishing contributions only in the distinguished basis $\textbf{r} \in \textbf{0} \cup \mathcal{N}$ it is sufficient to consider Eq.~\eqref{10112023_1057} in such a subspace only. The bare propagator reads
\begin{equation}
\label{03042024}
G_{0} = \hspace{-1.5mm}
\begin{pmatrix}
g_{00} &  g_{11} e^{F/2} & g_{01} &  g_{11} e^{-F/2} & g_{02} \\
g_{11}e^{-F/2} & g_{00} &  g_{10}e^{-F/2} &  g_{20}e^{-F} & g_{11}e^{-F/2} \\
g_{01} &  g_{10}e^{F/2} & g_{00} &  g_{10}e^{-F/2} & g_{01} \\
g_{11}e^{F/2} &  g_{20}e^{F} &  g_{10}e^{F/2} & g_{00} &  g_{11}e^{F/2} \\
g_{02} &  g_{11}e^{F/2} & g_{01} &  g_{11}e^{-F/2} & g_{00}
\end{pmatrix}.
\end{equation}
Some comments are in order. The quantity $g_{lm}$ denotes the propagator for $l$ steps along the $x$ direction and $m$ steps along the $y$ direction. The entries of this matrix can be readily obtained by referring to the distinguished basis illustrated in Fig.~\ref{fig_basis}. For the sake of clarity, we illustrate how to find them through some examples. The element $(1,3)$ represents the jump from $\textbf{e}_{0}$ to $- \textbf{e}_{y}$, therefore it corresponds to the propagator with $l=0$ and $m=1$, this yields $g_{01}$. Then, the entry $(4,1)$ corresponds to the jump from $- \textbf{e}_{y}$ to $\textbf{e}_{x}$, therefore it is associated with $l=1$ and $m=1$ and since the jump is parallel to the force, the propagator is multiplied by $\exp(F/2)$, for the reasons proved in Appendix \ref{Appendix_1}, this gives $g_{11}\exp(F/2)$. The \emph{free} propagators are defined via $g_{10}(s) := \langle \textbf{e}_x|\hat{G}_0( s^{\prime} )|\textbf{0}\rangle$, $g_{11}(s) := \langle \textbf{e}_x |\hat{G}_0( s^{\prime} )| \textbf{e}_y \rangle$, $g_{01}(s) := \langle \textbf{0} |\hat{G}_0( s^{\prime} )| \textbf{e}_y \rangle$, $g_{02}(s):= \langle \textbf{0} |\hat{G}_0( s^{\prime} )| 2\textbf{e}_y\rangle = \langle \textbf{e}_y|\hat{G}_0( s^{\prime} )|\!-\!\textbf{e}_y\rangle$. Note that the entries of the matrix $G_{0}$ are the propagators in the presence of the external force $F$. As we showed in Appendix \ref{Appendix_1}, the presence of the bias amounts to the factor $e^{\pm F/2}$ in front of the free propagator which is computed at the shifted frequency $s^{\prime} \equiv s+\Gamma-1$. Therefore, e.g., $g_{00}$ in [Eq.~(\ref{03042024})] stands for $g_{00}(s^{\prime})$. The quasi-confinement along $y$ implies that the transition amplitudes for $n$ steps along $x$ differ from the corresponding one for $n$ steps along $y$, e.g., $g_{10} \neq g_{01}$ and the equality is restored only in the limit $L \rightarrow \infty$. 

The solution of the scattering problem [Eq.~(\ref{10112023_1057})] amounts to solve a $5\times5$ matrix problem, something that can be conveniently pursued with the aid of symbolic computation programs. At a formal level, the solution is
\begin{equation}
\label{ }
t = S v \, ,
\end{equation}
where the $S$-matrix is
\begin{equation}
\label{ }
S = (1-v G_{0})^{-1} \, .
\end{equation}
Although this procedure is mathematically well-posed, the actual expression for the matrix $t$ turns out to be rather cumbersome. In Appendix \ref{sec_symmetry} we show how to implement the symmetries of the problem in order to reduce the computational complexity involved in the required matrix calculations.

%%%%%%%%%%%%%%%%%%%%%%%%%%%%%%%%%%%%%%%%%%%%%%%%%%%%%%%%%%%%%%%%%%%%%
%%%%%%%%%%%%%%%%%%%%%%%%%%%%%%%%%%%%%%%%%%%%%%%%%%%%%%%%%%%%%%%%%%%%%
\section{Velocity relaxation}
\label{sec_velocity_relaxation}
%%%%%%%%%%%%%%%%%%%%%%%%%%%%%%%%%%%%%%%%%%%%%%%%%%%%%%%%%%%%%%%%%%%%%
%%%%%%%%%%%%%%%%%%%%%%%%%%%%%%%%%%%%%%%%%%%%%%%%%%%%%%%%%%%%%%%%%%%%%
The time-dependent velocity response can be obtained from the characteristic function via $\img \partial^{2} F(\textbf{k},t)/\partial t \partial k_{x} \vert_{\textbf{k}=0}$; the corresponding expression in Laplace domain reads $\img s \partial [ G ]_{\rm av}(\textbf{k}) \partial_{k_{x}} \vert_{\textbf{k}=0}$. More explicitly, to first order in the obstacle density, we obtain
\begin{equation}
\label{05032024_1513}
\widehat{v}(s) = \frac{v_{0}}{s} + n \frac{v_{0}}{s} + n \frac{ \img  }{ s } \frac{ N \partial t(\textbf{k}) }{ \partial k_{x} } \bigg\vert_{\textbf{k}=0} \, ,
\end{equation}
where $v_{0}$ is the drift velocity in the absence of obstacles. The velocity at short times can be inferred from the high-frequency behavior since $v(t \rightarrow0) = \lim_{s \rightarrow \infty} s \widehat{v}(s)=(1-n)v_{0}$; this result can be interpreted as the bare velocity corrected by the fraction of accessible sites. This outcome follows because in the limit $s \rightarrow \infty$ the third term in [Eq.~(\ref{05032024_1513})] becomes twice the negative of the second. For long times a non-equilibrium stationary state is reached and the tracer attains the terminal velocity $v(t \rightarrow \infty)$, the latter is completely determined by the small-frequency behavior by means of the limit $v(t \rightarrow \infty) = \lim_{s \rightarrow0} s \widehat{v}(s)$. 

The obstacle-induced contribution to the velocity is completely encoded in the derivative of the forward scattering matrix in plane wave basis, which we write as follows
\begin{equation}
\begin{aligned}
\label{13102022_1103}
\img \partial_{k_{x}} N t(\textbf{k}) \big\vert_{\bf{k} = \bf{0}} & = - \sqrt{10} \langle p_{x} \vert \hat{t} \vert n \rangle \, .
\end{aligned}
\end{equation}

This result has been obtained by employing the symmetry-adapted basis outlined in Appendix \ref{sec_symmetry} which is inspired by the problem in absence of driving \cite{LSF_2018}. Although some of the symmetries are lost due to the driving, the residual symmetries are conveniently handled by the orthogonal transformation used in absence of driving. The right-hand side of Eq.~(\ref{13102022_1103}) is expressed in terms of a matrix element involving the forward scattering $t$-matrix. The vector $\vert n \rangle$ is related to conservation of probability via $\langle n \vert \hat{v}=0$, it is therefore a "neutral mode''. The vector $\vert p_{x} \rangle$ is reminiscent of a ``dipole'' oriented along the $x$-axis. In general, the matrix $t(\textbf{k})$ in plane wave basis can be decomposed into different channels corresponding to the aforementioned modes, as well as the dipole $\vert p_{y} \rangle$, the quadrupole $\vert d_{xy} \rangle$, and the $s$-wave $\vert s \rangle$. Rather elegantly, for the calculation of the velocity response the only non-vanishing contribution is the one that couples the neutral mode to the $p_{x}$-channel.

It turns out to be convenient to factor out the bare velocity by writing the left hand side as $\img \partial_{k_{x}} N t(\textbf{k}) \big\vert_{\bf{k} = \bf{0}} = v_{0} \mathcal{V}_{L}(F;s)$. By taking the low-frequency limit of Eq.~(\ref{05032024_1513}), we find the terminal velocity
\begin{equation}
\label{05032024_1425}
v(t \rightarrow \infty) = v_{0} + n v_{0} + n v_{0} \mathcal{V}_{L}(F;0) \, .
\end{equation}
The analytic expression of the function $\mathcal{V}_{L}(F;s)$ turns out to be rather cumbersome to be reported here; nevertheless, in Appendix \ref{two_lane} we show the explicit result for the case $L=2$. The procedure to obtain $\mathcal{V}_{L}(F;s)$ is illustrated in Appendix \ref{sec_moments} and explicit results for this function are available for any $L$ in the supporting \texttt{Mathematica} script~\cite{LorentzGasNotebook}.
%\red{online repository available at [...].} 

\begin{figure}[htbp]
\centering
\includegraphics[width=\columnwidth]{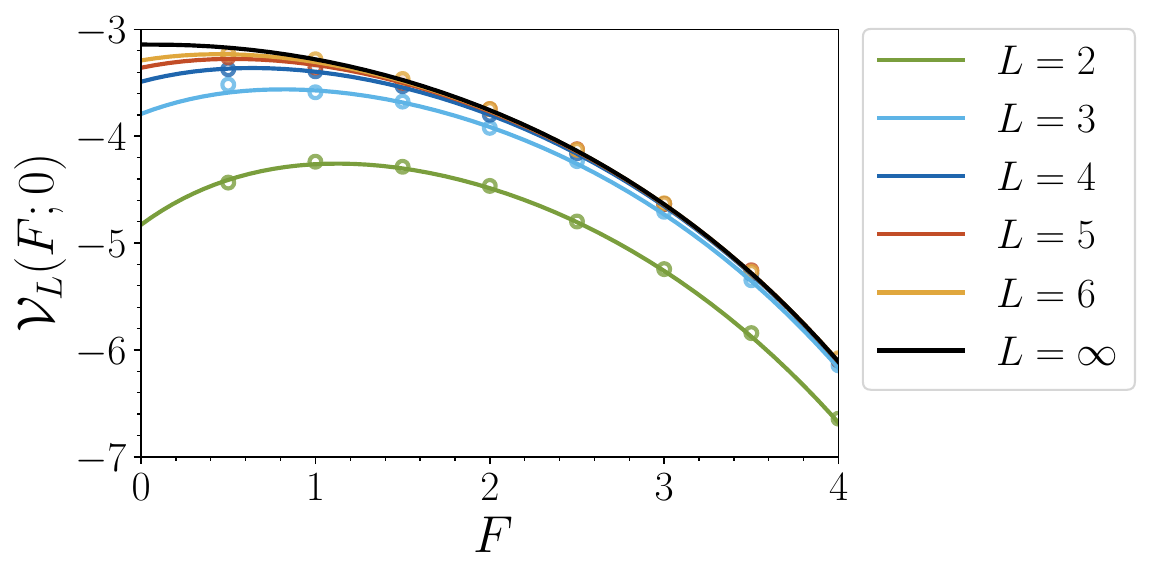}
\caption{The function $\mathcal{V}_{L}(F;0)$ as a function of $F$ for some values of the quasi-confinement size $L$. Symbols show the results obtained from stochastic simulations.}
\label{fig_function_V}
\end{figure}

 A plot of $\mathcal{V}_{L}(F;0)$ as a function of $F$ is provided in Fig.~\ref{fig_function_V}.

\begin{figure}[htbp]
\centering
\includegraphics[width=\columnwidth]{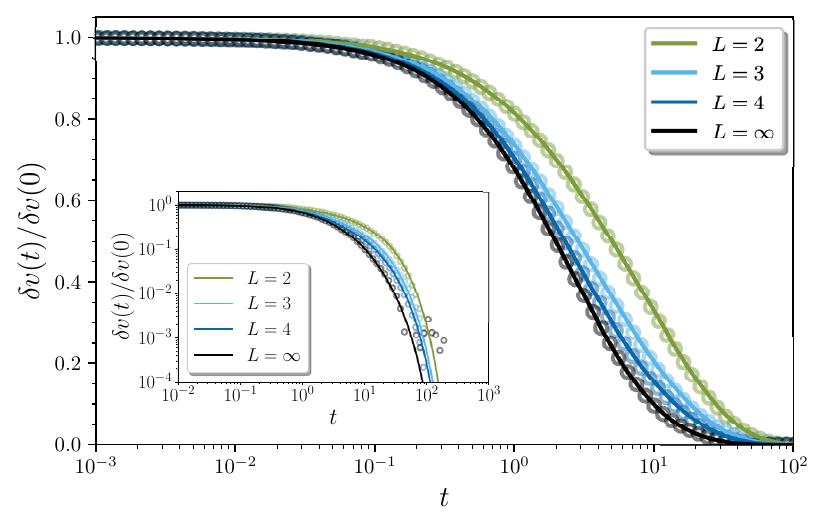}
\caption{Time-dependent approach to the stationary velocity $v(t)-v(t\rightarrow \infty)$ normalized to its initial value. Symbols correspond to simulations with $n=10^{-3}$ and $F=1$, solid lines represent the theoretical predictions. The inset represents the same quantity on logarithmic scales.}
\label{fig2}
\end{figure}

The velocity in Laplace domain can be written as $\widehat{v}(s) = v_{0}/s + n \Delta \widehat{v}(s)$, where the response term is $\Delta \widehat{v}(s) = [v_{0} + v_{0} \mathcal{V}_{L}(F;s)]/s$. The velocity in the time domain can be obtained by inverting the Laplace transform, therefore we can write $v(t) = v_{0} + n \Delta v(t)$, where $\Delta v(t)$ is the inverse Laplace transform of $\Delta \widehat{v}(s)$. The inversion is performed numerically by means of a suitable Filon formula \cite{Tuck}. An effective way to analyze the approach of the velocity $v(t)$ to its terminal value is to consider the normalized velocity response defined as $\delta v(t) = v(t)-v(t\rightarrow \infty)$ rescaled to its initial value $\delta v(0) = v(0)-v(t\rightarrow \infty)$. In Fig.~\ref{fig2}, we successfully tested the normalized velocity response $\delta v(t)/\delta v(0)$ for several values of the confinement size $L$ obtained from both theory and simulations for the intermediate value $F=1.0$. We observe that the increase of the confinement size yields a faster relaxation of the velocity to its final value. As a result, a strong confinement -- i.e., a small $L$ --  is accompanied by a slow relaxation, while for the unbounded model ($L=\infty$) the relaxation is the fastest possible, as illustrated in Fig.~\ref{fig2}.

\begin{figure}[htbp]
\centering
\includegraphics[width=\columnwidth]{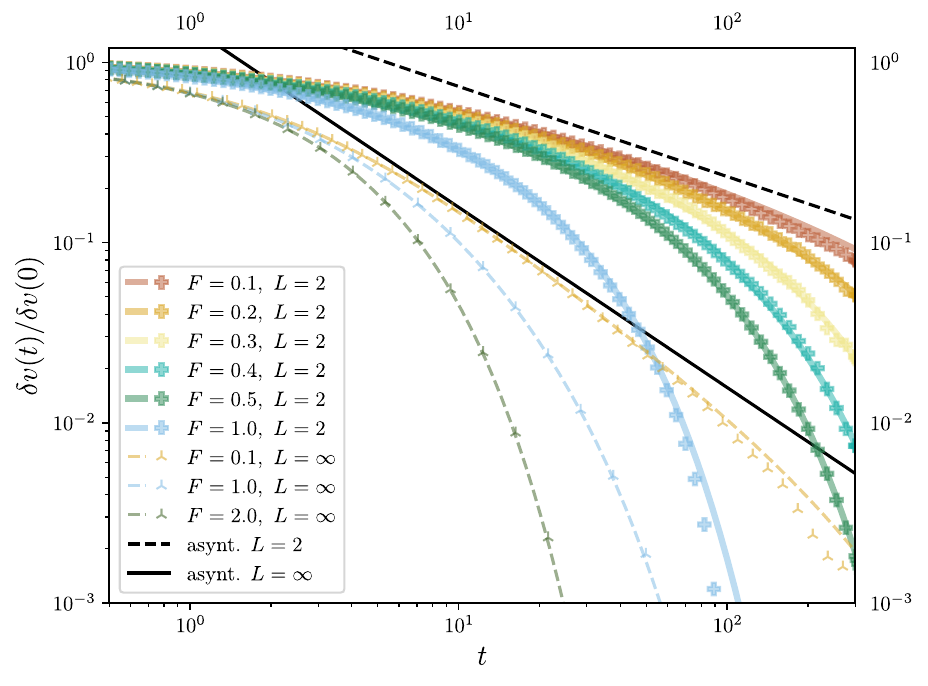}
\caption{Time-dependent approach to the stationary velocity $v(t)-v(t\rightarrow \infty)$ normalized to its initial value. Symbols correspond to simulations with $n=10^{-3}$. Solid and dashed lines represent the theoretical predictions for $L=2$ and $L=\infty$, respectively.}
\label{fig_relaxation}
\end{figure}

It is meaningful to ask what is the role played by the strength of the driving for a certain confinement $L$. This is addressed in the plot of Fig.~\ref{fig_relaxation} in which the normalized velocity response is plotted in double-logarithmic scale for several values of the force at $L=2$ and, for comparison, also $L=\infty$. Quite interestingly, as $F \rightarrow 0$ the velocity approaches the asymptotic value with a power-law of the form $t^{-1/2}$, this is shown in Fig.~\ref{fig_relaxation} with a dashed black line for $L=2$, however this feature persists for any finite $L$. For comparison we also illustrate the relaxation for the unbounded model, in this case the power-law is characterized by the power law $t^{-1}$, as shown with the solid black line in Fig.~\ref{fig_relaxation}.

These different exponents can be obtained as follows: within linear response -- legitimate for the regime $F \rightarrow 0$ -- the tail $t^{-(d+2)/2}$ is inherited from the long-time decay of the VACF; in particular, the velocity approach is given by $\delta v(t) = -F \int_{t}^{\infty}Z(t^{\prime}) \textrm{d}t^{\prime}$. From Fig.~\ref{fig_relaxation} one can infer that even for a very small driving the relaxation is no longer a power-law. In fact, the power-law tail is termed ``fragile'' in the sense that an infinitesimally small driving entails a decoration of the power-law with an exponential decay in the external force, as shown in Fig.~\ref{fig_relaxation} and in the inset of Fig.~\ref{fig2}. Hence, the drift $v(t)$ approaches the stationary state $v(t \rightarrow \infty)$ exponentially fast even for an arbitrarily small but finite driving force $F$, a result in strong contradistinction to the FDT predicting a slow algebraic approach.

\begin{comment}
\begin{eqnarray} \nonumber
v(s) & = & \img s \partial_{k_{x}} G_{0}(\textbf{k} \vert_{\textbf{k}=\textbf{0}} \\ \nonumber
& = & \img s^{-1} \partial_{k_{x}}\epsilon(\textbf{k}) \vert_{\textbf{k}=\textbf{0}} \\
& = & \frac{\sinh(F/2) }{2s} \, ,
\end{eqnarray}
in the last line we used $\epsilon(\textbf{0})=0$ and 
\begin{eqnarray} \nonumber
\partial_{k_{x}}\epsilon(\textbf{k}) \vert_{\textbf{k}=\textbf{0}} & = & - \img\Gamma \bigl[ W(\textbf{e}_{x}) - W(-\textbf{e}_{x}) \bigr] \\ \nonumber
& = & -\frac{\img}{2} \sinh(F/2) \, .
\end{eqnarray}
\end{comment}

%%%%%%%%%%%%%%%%%%%%%%%%%%%%%%%%%%%%%%%%%%%%%%%%%%%%%%%%%%%%%%%%%%%%%
%%%%%%%%%%%%%%%%%%%%%%%%%%%%%%%%%%%%%%%%%%%%%%%%%%%%%%%%%%%%%%%%%%%%%
\section{Fluctuations along the force}
\label{sec_fluctuations}
In the presence of an external force the system relaxes towards a non-equilibrium stationary state. The latter is characterized by a terminal velocity which has been obtained in Sec.~\ref{sec_velocity_relaxation} by analyzing the first moment of the displacement along the force. Higher-order moments -- all of them obtainable with the scattering approach -- provide a valuable tool for the characterization of fluctuations in the non-equilibrium stationary state. This section presents the analysis of fluctuations captured by the second moment in the direction of the force and their interplay with driving and confinement. The quantity of interest is the variance of the displacement
\begin{equation}
\label{ }
\textrm{Var}(t) = \langle \Delta x(t)^{2} \rangle -  \langle \Delta x(t) \rangle^{2} \, ,
\end{equation}
where $\langle \Delta x(t)^{2} \rangle$ is the mean-square displacement along the force. Other quantities that we will consider are the time-dependent diffusion coefficient,
\begin{align}
\label{29022024_1453}
D(t) := \frac{1}{2}\frac{\diff}{\diff t} \textrm{Var}(t) \, ,
\end{align}
the local exponent $\alpha(t)$ of the variance defined via
\begin{equation}
\label{ }
\alpha(t) := \frac{ \textrm{d} \ln \textrm{Var}(t) }{ \textrm{d} \ln t} = \frac{ 2 D(t) t }{ \textrm{Var}(t) } \, ,
\end{equation}
and the velocity autocorrelation function,
\begin{equation}
\label{27022024_1218}
Z(t) := \frac{1}{2} \frac{\textrm{d}^{2}}{\textrm{d}t^{2}}\textrm{Var}(t) \, .
\end{equation}
In Sec.~\ref{sec_vacf} we will examine $Z(t)$ and the long-time diffusion coefficient $D(t \rightarrow \infty)$ in equilibrium. The analysis of the fluctuations along the force for the non-equilibrium setup ($F>0$) as well as the force-induced diffusion coefficient are discussed in Sec.~\ref{sec_msd}.

%%%%%%%%%%%%%%%%%%%%%%%%%%%%%%%%%%%%%%%%%%%%%%%%%%%%%%%%%%%%%%%%%%%%%
%%%%%%%%%%%%%%%%%%%%%%%%%%%%%%%%%%%%%%%%%%%%%%%%%%%%%%%%%%%%%%%%%%%%%
\subsection{Velocity autocorrelation function and diffusion coefficient at equilibrium}
\label{sec_vacf}
To begin, we discuss the fluctuations in the absence of the external force. From Eqs.~(\ref{29022024_1453}) and (\ref{27022024_1218}) the VACF can be equivalently defined in term of the time-dependent diffusion coefficient by
\begin{equation}
\label{29022024_1451}
Z(t) = \frac{ \textrm{d} D(t) }{ \textrm{d} t } \, .
\end{equation}
It is convenient to pass to the frequency domain by using the Laplace transform defined via $\widehat{D}(s) := \mathcal{L}\{ D(t) \}(s) = \int_0^\infty e^{-s t} D(t) \diff t $. Once Eq.~(\ref{29022024_1451}) is translated in frequency domain, we have $\widehat{Z}(s)=s\widehat{D}(s)$. In order to find $Z(t)$ we first compute $D(t)$ by inverting the Laplace transform $\widehat{D}(s)$ and finally we perform the time derivative to find the VACF through Eq.~(\ref{29022024_1451}).

The time-dependent diffusion coefficient in the frequency domain follows by the rules of the Laplace transform,
\begin{equation}
\label{ }
\widehat{D}(s) = (s/2)\mathcal{L}\{ \langle \Delta x(t)^2 \rangle\}(s) \, .
\end{equation}
The mean-square displaced is obtained by taking successive $k_{x}$-derivatives of the disorder-averaged propagator, which in frequency domain becomes
\begin{align}
\label{}\nonumber
& \mathcal{L}\{ \langle \Delta x(t)^2 \rangle\}(s) = - \left. \frac{\partial^2}{\partial k_x^2} [ G _{\text{av}}]^c(\textbf{k}) ]  \right|_{\textbf{k}=0} \\
& = - \left. \frac{\partial^2 G_0}{\partial k_x^2} \right|_{\textbf{k}=0} + n \left[ -  \frac{\partial^2 G_0}{\partial k_x^2}  - G_0^2 \frac{\partial^2 N t(\textbf{k})}{\partial k_x^2}\right]_{\textbf{k}=0} \, .
\end{align}
We stress that for the unbiased motion ($F=0$) there is no average motion of the tracer along the unbounded direction. Therefore the first non-trivial cumulant is given by the mean-square displacement $\langle \Delta x(t)^2 \rangle$. The result simplifies to
\begin{equation}
\label{13062022_1351}
\begin{aligned}
\frac{\partial^2 N t(\textbf{k})}{\partial k_x^2} \biggr|_{\textbf{k}=0} & = 4 \langle p_{x} \vert \hat{t} \vert p_{x} \rangle + \sqrt{5} \langle d_{xy} \vert \hat{t} \vert n \rangle - \langle s \vert \hat{t} \vert n \rangle \, ,
\end{aligned}
\end{equation}
however, in equilibrium only the first term contributes, this is given by $\langle p_x | \hat{t} | p_x \rangle = 1/\Delta_{L}(s)$ and $\Delta_{L}(s) := 4- g_{00}(s) + g_{20}(s)$. Again, by using the symmetry-adapted basis a major simplification occurs: the only non-vanishing matrix element is the ones that couples the dipole $p_{x}$ to itself while the other channels do not contribute; we refer to Appendix \ref{sec_symmetry} for details. Therefore,
\begin{align}
\label{10112023_1115}
\widehat{Z}_{L}(s) = & \frac{1}{4} + n \left[ \frac{1}{4} -  \frac{ 2 }{ \Delta_{L}(s) } \right] \, .
\end{align}
Here we quote the explicit expressions for the combination of propagators entering in the response contribution to the VCAF; for the few lowest values of $L$ we obtain:
\begin{eqnarray}
\label{11042024_1435}
\nonumber
\Delta_{2}(s) & = & 4 \left(-2 s+\sqrt{s (s+1)}+\sqrt{(s+1) (s+2)}-1\right) \\ \nonumber
\Delta_{3}(s) & = & \frac{4}{3} \left(-6 s+2 \sqrt{s (s+1)}+\sqrt{(4 s+3) (4 s+7)}-3\right) \\ \nonumber
\Delta_{4}(s) & = & 2 \Bigl( -4 s+\sqrt{s (s+1)}+\sqrt{(s+1) (s+2)} \\
& + & \sqrt{4 s (s+2)+3}-2 \Bigr) \, .
\end{eqnarray}

It is useful to expand $\Delta_{L}(s)$ around $s=0$; leaving the details of the derivation in Appendix \ref{sec_low_freq_quasi_confined}, the expansion reads
\begin{equation}
\label{25092023_1114}
\Delta_{L}(s) = C_{L} +  8\sqrt{s}/L + O(s) \, , \qquad s \rightarrow 0 \, ,
\end{equation}
where $C_{L}$ is a confinement-dependent constant
\begin{equation}
\label{22092023_1208}
C_{L} = -4 + \frac{4}{L} \sum_{q=1}^{L-1} \sqrt{(2-\cos(2\pi q/L))^{2}-1} \, .
\end{equation}
The diffusion coefficient at infinite times can be obtained by means of the limit $D_{x}^{\rm eq} := \lim_{t \rightarrow \infty} D_{x}(t) = \lim_{s \rightarrow 0} s \widehat{D}(s) = \widehat{Z}_{L}(0)$; this result can be readily obtained by virtue of Eq.~(\ref{25092023_1114}), hence we obtain
\begin{equation}
\label{27022024_1226}
D_{x}^{\rm eq} = \frac{1}{4} + n \left( \frac{1}{4} - \frac{2}{C_{L}} \right) \, .
\end{equation}
The small-frequency expansion [Eq.~(\ref{25092023_1114})] implies that for $s \rightarrow 0$
\begin{equation}
\label{ }
\widehat{Z}_{L}(s) =  \frac{1}{4} + n \left[ \frac{1}{4} - \frac{2}{C_L}  + \frac{16}{L C_L^2 }\sqrt{s} + O(s) \right] \, .
\end{equation}
The leading non-analytic contribution of order $O(s^{1/2})$ drops out in the limit $L\to \infty$ since the particle is no longer confined and, as expected, the VACF reduces to
\begin{align} 
\widehat{Z}_{L}(s) = \frac{1}{4} - n \frac{\pi-1}{4} + O(s) \, ,
\end{align}
a result that follows from $\Delta_{\infty}(0)=C_{\infty}=8/\pi$ [Eq.~(\ref{29022024_1513})]. By taking $s \rightarrow 0$ in the above we retrieve the well-known diffusion coefficient to order $n$ for the unbounded model \cite{OK_nieuwenhuizen_3}.

\begin{figure}[htbp]
\centering
\includegraphics[width=\columnwidth]{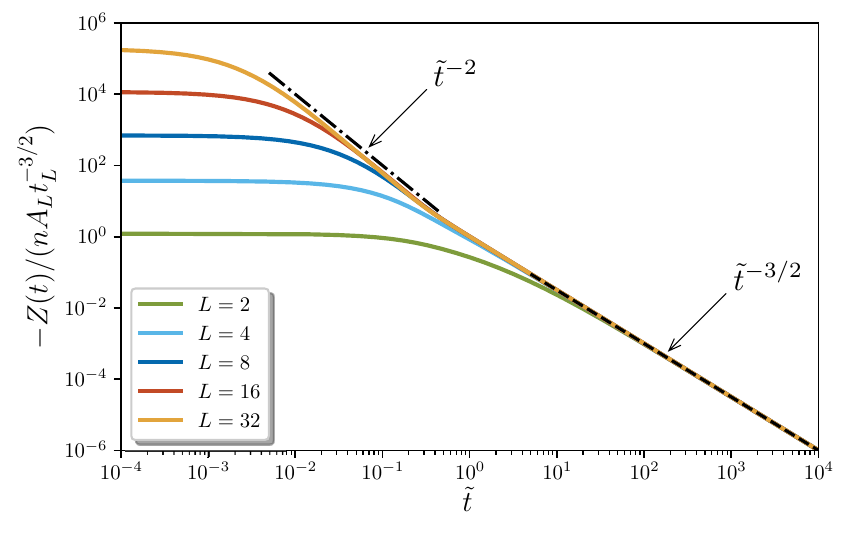}
\caption{The VACF rescaled to its asymptotic tail $- n A_{L} t_{L}^{-3/2}$. The dashed black line indicates the exact long-time tail $\propto \tilde{t}^{-3/2}$, the dot-dashed corresponds to the transient tail $\propto \tilde{t}^{-2}$.
}
\label{fig_vacf_master}
\end{figure}

We are now in the position to discuss the VACF in the time domain. To this end, we briefly recall how to perform the inverse of the Laplace transform. To be definite let us consider the Laplace transform of a function $f(t)$, which is denoted $\widehat{f}(s) \equiv \int_{0}^{\infty}\textrm{d}t \, \textrm{e}^{-st} f(t)$. If $\lim_{t \rightarrow \infty} f(t)=0$, then the function $f(t)$ can be obtained by means of the integral along the imaginary axis in the frequency domain by means of \cite{LSF_2018}
\begin{equation}
\label{05032024_1427}
f(t) = \frac{2}{\pi} \int_{0}^{\infty}\textrm{d}\omega \, \textrm{Re}\bigl[ \widehat{f}(\img \omega) \bigr] \cos(\omega t) \, .
\end{equation}
The above computation scheme cannot be directly applied to $\widehat{D}(s)$ because $D(t)$ admits a finite nonzero value $D_{x}^{\rm eq}$ at infinite times. In order to apply the above procedure we decompose $\widehat{D}(s)$ by writing
\begin{equation}
\label{ }
\widehat{D}(s) = \frac{ D_{x}^{\rm eq} }{s} + \left( \widehat{D}(s) - \frac{ D_{x}^{\rm eq} }{s} \right) \, ,
\end{equation}
where $\widehat{D}(s)$ is obtained from scattering theory [Eq.~(\ref{10112023_1115})]. The inverse of the Laplace transform of the first term is trivial since it yields the behavior $D_{x}^{\rm eq}t$, while for the term in brackets we are allowed to apply the scheme mentioned above. Therefore the time-dependent diffusion coefficient is
\begin{equation}
\label{ }
D(t) = D_{x}^{\rm eq} + \frac{2}{\pi} \int_{0}^{\infty}\textrm{d}\omega \, \textrm{Re}\biggl[ \widehat{D}(\img \omega) - \frac{ D_{x}^{\rm eq} }{\img \omega} \biggr] \cos(\omega t) \, .
\end{equation}
By performing a derivative with respect to the time we obtain the VACF in time domain, thus a simple rearrangement yields
\begin{equation}
\label{25092023_1255}
Z_{L}(t) = \frac{4n}{\pi} \int_{0}^{\infty} \textrm{d}\omega \, \textrm{Im}\biggl[ \frac{1}{\Delta_{L}(\img\omega)} \biggr] \sin(\omega t) \, .
\end{equation}
The converge of the integral is assured since the integrand behaves as $\omega^{-1}$ for $\omega \rightarrow \infty$ and $\sin(\omega t)/\omega$ is integrable. From the low-frequency behavior [Eq.~(\ref{25092023_1114})] we infer that the VACF exhibits a long-time tail of the form
\begin{equation}
\label{10112023_1128}
Z_{L}(t) \sim - A_{L} n t^{-3/2} \, , \qquad t \rightarrow \infty \, ,
\end{equation}
where $A_{L}$ is the $L$-dependent prefactor
\begin{equation}
\label{ }
A_{L} = \frac{8}{ \sqrt{\pi} LC_{L}^{2}} \, .
\end{equation}
For any finite $L$ the prefactor $A_{L}$ is finite while for $L \rightarrow \infty$ we have $A_{L} \sim p/L$ with coefficient $p=\pi^{3/2}/8$. For the sake of completeness, the occurrence of the long-time tail $\propto t^{-3/2}$ has been reported for a one-dimensional Lorentz-type model \cite{Grassberger_1980} although in such a model the VACF exhibits an oscillatory behavior. In the accompanying Letter we show that for large but finite $L$ there exists a transient time domain in which the decay is of the form $\propto t^{-2}$ and that the long-time tail $\propto t^{-3/2}$ is ultimately attained for $t \gg t_{L}$ where $t_{L} \propto L^{2}/D$ is the time scale associated to the exploration of the periodic direction. Hence, we plot the VACF rescaled to its asymptotic tail [Eq.~(\ref{10112023_1128})] as a function of the rescaled time $\tilde{t} : = t/t_{L}$; without loss of generality we can take $t_{L} \equiv L^{2}$ and the result we obtain is the shown in Fig.~\ref{fig_vacf_master}. We observe that the data collapse to the master curve protracts from large scaled times $\tilde{t} \gg 1$ down to small times $\tilde{t} \ll 1 $ upon increasing $L$. We further note that the scaling breaks down for microscopic times $t = O(\tau)$; in terms of rescaled times this implies a behavior proportional to $\tilde{t}^{-2}$ for $\tilde{t} \gg \tau/t_{L}$ and $\tilde{t}^{-3/2}$ for $\tilde{t} \gg 1$. In the limit $L \rightarrow \infty$ and $t \rightarrow \infty$ with fixed $t/t_{L}$ the data exhibit a collapse onto a single master curve whose behavior at large times is $\tilde{t}^{-3/2}$ while for short times it behaves as $\pi^{-1/2}\tilde{t}^{-2}$, these asymptotic power laws are indicated with dashed and dot-dashed black curves in Fig.~\ref{fig_vacf_master}.

The black curve in Fig.~\ref{fig_vacf_master} is obtained by inverting the Laplace transform $\widehat{Z}_{L}(s)$ for the unbounded model. The latter is formally obtained by taking the limit $L \rightarrow \infty$ in $\widehat{Z}_{L}(s) = [ 1 + n (1+\mathcal{V}_{L}(0,s))]/4$, where $\mathcal{V}_{L}(0;s)=-8/\Delta_{L}(s)$ is the function already discussed in the context of the velocity relaxation, however, in this case it is evaluated for $F=0$ and $s \neq 0$. As anticipated, the limit can be performed at a formal level while its implementation is far from being trivial since propagators themselves are expressed as a sum over modes $k_{y}=2\pi q/L$ labelled by an integer running from $q=1$ to $q=L-1$; we refer to Appendix \ref{Appendix_1} for further details on lattice Green's functions. Nonetheless, results for $L \rightarrow \infty$ are more conveniently retrieved simply by examining the solution of the unbounded model which yields \cite{OK_nieuwenhuizen_3, LF_2013}, in our notations,
\begin{equation}
\label{03042024_1513}
\mathcal{V}_{\infty}(0;s) = \frac{2\pi}{\pi s -2(1+s)\textsf{E}[(1+s)^{-2}]} \, ,
\end{equation}
where $\textsf{E}[\kappa] = \int_{0}^{\pi/2}\textrm{d}\theta \left( 1-\kappa\sin^{2}\theta \right)^{1/2}$ is the complete elliptic integral of the second kind \cite{NIST, DLMF}; in this paper we adopted \texttt{Mathematica}'s convention for elliptic integrals. For the sake of completeness we also provide the quantity $\Delta_{L}(s)$ for the unbounded model. The  matrix element of the forward-scattering matrix can expressed in terms of an elliptic integral and the result takes the form
\begin{eqnarray} \nonumber
\label{25092023_1254}
\Delta_{\infty}(s) & = & -4s + \frac{8}{\pi} (1+s)\textsf{E}[(1+s)^{-2}] \\ \nonumber
& = & \frac{8}{\pi} + \biggl[ - 4 + \frac{4}{\pi} + \frac{12}{\pi}\ln2 \biggr] s - \frac{4}{\pi}s\ln s + O(s^{2}) \, ,
\end{eqnarray}
the last line holds as $s \rightarrow 0$ and is useful for the analysis of the long-time tail. The small-frequency expansion of the VACF contains smooth terms in $s$ and the first non-regular term in such an expansion is proportional to $\propto s\ln s$ with prefactor $-\pi/8$. The latter yields the leading form of the long-time tail which, including the prefactor, is
\begin{equation}
\label{ltt_non_confined}
Z_{\infty}(t) \sim - \frac{\pi n}{8} t^{-2} \, , \qquad t \rightarrow \infty \, ,
\end{equation}
a result that can be obtained by virtue of Tauber's theorem (see, e.g., \cite{Feller}).

In Appendix \ref{scaling_diffusion_coefficient} we extend the above analysis to the time-dependent diffusion coefficient and its scaling form analogous to the one shown in Fig.~\ref{fig_vacf_master} for the VACF.

%%%%%%%%%%%%%%%%%%%%%%%%%%%%%%%%%%%%%%%%%%%%%%%%%%%%%%%%%%%%%%%%%%%%%
%%%%%%%%%%%%%%%%%%%%%%%%%%%%%%%%%%%%%%%%%%%%%%%%%%%%%%%%%%%%%%%%%%%%%
\subsection{Diffusion coefficient and mean-square displacement away from equilibrium}
\label{sec_msd}
\begin{figure}[htbp]
\centering
\includegraphics[width=\columnwidth]{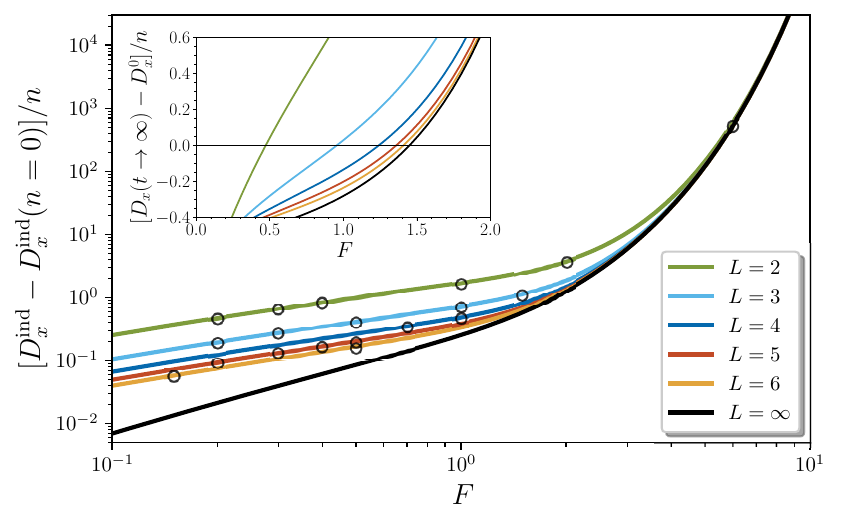}
\caption{Force-induced diffusion coefficient $D_{x}^{\rm ind}$ [corrected by the empty lattice $D_{x}^{\rm ind}(n=0)$] in units of the bare diffusion coefficient $D_{x}^{0}(F)$. Results from simulations are indicated with circles, the analytic result correspond to solid lines.}
\label{Fig_5_dind}
\end{figure}

We can now present the results obtained for the non-equilibrium setup in which the pulling force $F$ can be arbitrarily strong. The first quantity we discuss is the long-time diffusion coefficient and how it is affected by the size $L$ of the quasi-confinement. A convenient way to rationalize the effects due to the force is to consider the \emph{force-induced diffusion coefficient}
\begin{equation}
\label{20092023_1428}
D_{x}^{\rm ind} = D_{x}(t \rightarrow \infty) - D_{x}^{\rm eq}(t \rightarrow \infty) \, ,
\end{equation}
where $D_{x}^{\rm eq}(t \rightarrow \infty)$ is the $L$-dependent long-time diffusion coefficient in the absence of driving [Eq.~(\ref{27022024_1226})]; clearly, the force-induced diffusion coefficient depends on the obstacle density, force and confinement, thus, $D_{x}^{\rm ind} \equiv D_{x}^{\rm ind}(F,n,L)$. The theory gives for the infinite-time diffusion coefficient
\begin{equation}
\label{ }
D_{x}(t \rightarrow \infty) : = D_{x}^{0}(F) + n \Xi_{L}(F) + O(n^2) \, ,
\end{equation}
where $\Xi_{L}(F)$ is known exactly for any system size $L$ and arbitrary force $F$. We emphasize that the force-dependent first-order correction $\Xi_{L}(F)$ is formally defined as
\begin{equation}
\label{ }
\Xi_{L}(F) = \lim_{ n \rightarrow 0} \frac{D_{x}(t \rightarrow \infty) - D_{x}^{0}(F)}{n} \, ,
\end{equation}
where $D_{x}^{0}(F)$ is the bare diffusion coefficient [Eq.~\eqref{09122025_1040}] and the infinite-time diffusion coefficient is found from its Laplace transform, $D_{x}(t \rightarrow \infty) = \lim_{s \rightarrow 0} s \widehat{D}(s)$, with $\widehat{D}(s)$ is obtained from the scattering formalism, as explained in Appendix~\ref{Appendix_2}. Explicit analytic results for $\Xi_{L}(F)$ are available in the supporting \texttt{Mathematica} script~\cite{LorentzGasNotebook}. In particular, for $F=0$ one finds $\Xi_{L}(0) = (1/4) - 2/C_{L}$ and the corresponding result in equilibrium [Eq.~(\ref{27022024_1226})] is retrieved. In order to single out the effects due to the disorder we consider the force-induced diffusion coefficient corrected by the empty lattice ($n=0$), this subtraction entails
\begin{equation}
\label{20092023_1541}
D_{x}^{\rm ind} - D_{x}^{\rm ind}(F,n=0,L) = n \bigl[ \Xi_{L}(F)-\Xi_{L}(0) \bigr] \, .
\end{equation}
The above quantity is extracted in simulations and compared to theory in Fig.~\ref{Fig_5_dind}. 

The analytical solution shows that for any finite $L$ the corrected force-induced diffusion coefficient [Eq.~(\ref{20092023_1541})] vanishes as $b_{L}|F|$ for $F \rightarrow 0$. This linear behavior is in sharp contrast with the asymptotic behavior $\propto n F^{2} \ln|F|$ observed for the unbounded model \cite{LF_2017, LSF_2018}. The constant $b_{L}$ decreases as $L$ increases and it can be calculated for any $L$; here some values: $b_{2}=(12+7\sqrt{2})/8$, $b_{3}=\left(31+5\sqrt{21}\right)/48$. As a result of the confinement, the corrected force-induced diffusion coefficient for small driving is enhanced by orders of magnitude with respect to the unconfined model. A completely different scenario is observed for large forces $F \gtrsim 6$. Irrespectively of the size $L$ the force-induced diffusion coefficient retrieves the asymptotic expression for the unbounded model,
\begin{equation}
\label{ }
D_{x}^{\rm ind} = \frac{n}{16} \textrm{e}^{3F/2} + O(\textrm{e}^{F}) \, , \quad F \rightarrow \infty \, ,
\end{equation}
whose leading asymptotic behavior is indeed size-independent. Therefore, $L$-dependent corrections are expected to appear via corrections that are subdominant with respect to $\exp(3F/2)$ as $F \rightarrow \infty$.

Another intriguing feature of the model is the occurrence of a critical force $F_{c,L}$ such that for $F<F_{c,L}$ the increase of disorder is accompanied by a suppression of the long-time diffusion coefficient, i.e., $D_{x}(t \rightarrow \infty) < D_{x}^{0}$. Then, for $F>F_{c,L}$ increasing the obstacle density yields an enhancement of the diffusivity, as shown in the inset of Fig.~\ref{Fig_5_dind}. Some values of $F_{c,L}$ are tabulated in Tab.~\ref{tab}. The analytic solution of the model allows us to conclude that the above scenario actually occurs for any choice of the confinement size $L$. Upon decreasing size $L$ the critical force $F_{c,L}$ reduces from $F_{c,\infty} \approx 1.45$ to $F_{c,2} \approx 0.47$ and, as a result, the domain in which the disorder suppresses fluctuations shrinks. The decreasing of the threshold value $F_{c,L}$ means that confinement favors the onset of enhancement of the disorder-induced diffusivity. 
\begin{table}[ht]
\centering
\begin{tabular}[t]{lccc}
\toprule
$L$ & $F_{c,L}$ & $C_{L}$ & $C_{L} (\text{approx})$ \\
\midrule
2	& 0.4723 & $4  \sqrt{2}- 4 $ & $1.6569$ \\
%0.472297
3	& 0.9556 & $4 \sqrt{7/3}-4$ & $2.1101$ \\
%0.955603
4	& 1.2412 & $2\sqrt{2}+ 2\sqrt{3} -4$ & $2.2925$ \\
%1.24124
5	& 1.3597 & $\frac{4}{5} \sqrt{35  + 2 \sqrt{205}}-4$ & $2.3818$ \\
%1.35956
6	& 1.4088 & $\frac{2}{3} \left(2 \sqrt{2}+\sqrt{5}+\sqrt{21}\right)-4$ & $2.4314$ \\
$\infty$ & 1.4495 & $8/\pi$ & $2.5465$ \\
%1.44949
\bottomrule
\end{tabular}
\caption{The critical force $F_{c,L}$ and the constant $C_{L}$ for several values of the quasi-confinement size.}
\label{tab}
\end{table}%

The theory shows that for any value of the force strength and confinement the variance of the displacement grows linearly in time in the large-time regime. From the resulting prefactor we can determine the diffusion coefficient in the non-equilibrium stationary state. The actual value of the long-time diffusion coefficient can be orders of magnitude higher than the one in absence of driving. This picture persists in the presence of confinement although confinement-induced effects on the diffusion coefficient become minute at large forces. On the other hand, the confinement plays a major role at small forces. In particular, the force-induced diffusion coefficient increases upon confining. Furthermore, the nature of the non-analytic behavior at $F \rightarrow 0$ depends sensitively on the confinement.

\begin{figure}[htbp]
\centering
\includegraphics[width=\columnwidth]{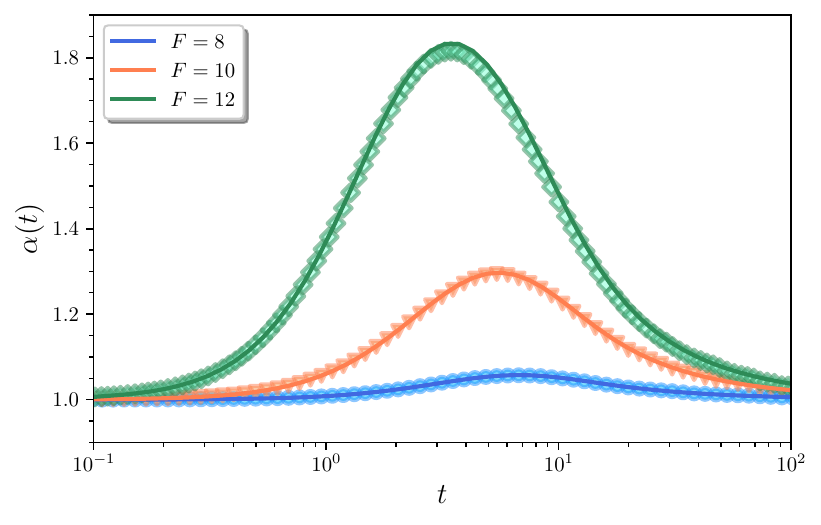}
\caption{Time-dependent local exponent of the variance $\alpha(t)$ along the force for $n=10^{-4}$. Simulation data are indicated with symbols, the theoretical prediction is indicated wit solid lines.}
\label{Fig_alpha}
\end{figure}

The complete time-dependent behavior of the variance can be accessed by means of the scattering formalism. In Fig.~\ref{Fig_alpha} we show the comparison between theory and simulations for the time-dependent behavior of the local exponent of the variance for the specific case $L=2$.
The results of Fig.~\ref{Fig_alpha} show the emergence of a transient superdiffusive behavior at intermediate times. For fixed time, the local exponent $\alpha(t)$ increases upon increasing the force. However, for forces larger than the ones shown in the figure the exponent saturates to $\alpha(t)=3$ as in the unbounded model \cite{LF_2017} but still transiently, i.e., the saturation occurs as a peak that spreads forming a plateau in a certain time-window. Further simulations we performed at $L=2$ indicate that upon lowering the obstacle density the exponent $\alpha(t)$ increases and the peak shifts to shorter times, in close analogy to the unbounded model. Then, for all values of the force the behavior $\alpha(t) =1$ is retrieved both at short and infinite times, as shown in Fig.~\ref{Fig_alpha}.

%%%%%%%%%%%%%%%%%%%%%%%%%%%%%%%%%%%%%%%%%%%%%%%%%%%%%%%%%%%%%%%%%%%%%
%%%%%%%%%%%%%%%%%%%%%%%%%%%%%%%%%%%%%%%%%%%%%%%%%%%%%%%%%%%%%%%%%%%%%
\section{Summary and conclusions}
\label{sec_summary_and_conclusions}
In this paper we have solved for the first time the complete time-dependent dynamics of a tracer particle exploring a random landscape of impurities in a quasi-confined channel under the action of an arbitrary strong driving force. After having mapped the master equation describing the site-occupation probability onto a Schr\"odinger equation, we introduced the obstacles on the lattice by modeling them as impenetrable sites. Then, by treating the obstacles as perturbations of a free Hamiltonian, we illustrated how to employ the quantum-mechanical scattering formalism in order to find the disordered-average propagator exact to first order in the obstacle density. This is achieved by solving the scattering problem for a single impurity, a $5 \times 5$ matrix problem that can be tackled with the aid of computer algebra.

We find that -- already in equilibrium -- the quasi-confined lattice Lorentz model in the absence of the external force displays the intriguing feature of a \emph{dimensional crossover}. We showed this property by an exact analysis of the equilibrium VACF. The latter is characterized by a transient long-time tail of the form $t^{-2}$ -- reminiscent of the unbounded model --, while the long-time asymptotic behavior is a power law $\propto t^{-3/2}$. These different behaviors can be rationalized in terms of a dimensional-dependent long-time tail of the form $t^{-(d+2)/2}$. The analytic solution shows that for any finite $L$ the effective dimensionality is $d=1$ at large times while $d=2$ at intermediate times. The two regimes are separated by a size-dependent time scale $t_{L} =L^{2}/D$ which can be interpreted as the time to explore the confined direction.

We then switched on the driving force. When the tracer particle is pulled a non-equilibrium stationary state is reached in which the average motion occurs with a force-dependent terminal velocity that we found analytically. The instantaneous velocity relaxes towards its terminal value via a power-law $t^{-(d+2)/2}$ decorated with an exponential time-dependence due to the force; the effective dimensionality is $d=1$ for any finite confinement $L$. As a result, the long-time tail is \emph{fragile}, as it happens for the unbounded model and in such a case $d=2$. We analyzed the fluctuations of the tracer in the direction parallel to the force. For large times the variance of the displacement along the force grows linearly with time and its prefactor is a force-dependent and confinement-dependent diffusion coefficient. For large forces the confinement plays a minor role, as expected on intuitive basis since the particle hops rarely in the confined direction. On the other hand the diffusion coefficient turns out to be sensitive to the confinement in the regime of small driving. In this regime the force-dependent diffusion coefficient of the unbounded model exhibits non-analytic behavior of the form $F^{2}\ln|F|$. The occurrence of non-analyticity persists upon confinement, now with a non-analytical behavior of the form $b_{L}|F|$, with an amplitude $b_{L}$ that we obtained analytically. Although it is expected the suppression of diffusivity upon increasing the disorder, we found that this is true only when the force is smaller than a critical force $F_{c,L}$. Quite interestingly, if $F>F_{c,L}$ then the increase of disorder leads to an enhanced diffusivity. We find that $F_{c,L}$ decreases upon increasing $L$, meaning that the onset for the disorder-enhancement diffusion regime is favored by the confinement. Finally, the complete time-dependent solution shows that the behaviors at short and large times are interpolated by a transient superdiffusive behavior. We characterized this feature in terms of the local exponent of anomalous diffusion $\alpha(t)$ by showing how it increases upon increasing of the force. In this regard, especially for large forces, the confined model behaves analogously to the unbounded one and therefore the saturation to $\alpha(t)=3$ is expected as a transient phenomenon followed by diffusion at infinite times.

An important point should be made explicit. For any finite obstacle density $n>0$, there is a nonzero probability that the channel contains a blocking cluster of obstacles. As the tracer explores longer and longer distances, the probability of encountering such a cluster approaches unity; hence, at sufficiently long times the tracer becomes immobilized. Consequently, in the infinite-time limit, transport coefficients such as the diffusion coefficient vanish for any finite $n$.
This implies the existence of a clogging time scale $t_*(n,F)$ (depending on $n$ and possibly on the drive $F$) beyond which clogging dominates. Our numerical simulations probe the pre-asymptotic regime $t\ll t_*$; within this time window the tracer remains mobile and no clogging is observed.

To conclude, we showed how to solve exactly the lattice Lorentz model in the presence of quasi-confinement and how this model provides highly non-trivial insights on the behavior of a strongly interacting system out of equilibrium. Looking at future perspectives, it would be interesting to investigate the quasi-confinement in a three-dimensional system with periodicity in two directions. This is actually possible by amalgamating the techniques presented in this paper for the quasi-confinement together with the solution of the unbounded three-dimensional model \cite{LBF_2018}. In such a case, a richer variety of dimensional crossovers is expected; we plan to address this issue in the near future.

%%%%%%%%%%%%%%%%%%%%%%%%%%%%%%%%%%%%%%%%%%%%%%%%%%%%%%%%%%%%%%%%%%%%L%
%%%%%%%%%%%%%%%%%%%%%%%%%%%%%%%%%%%%%%%%%%%%%%%%%%%%%%%%%%%%%%%%%%%%%
\section*{Acknowledgments}
 This research was funded in
part by the Austrian Science Fund (FWF) 
through the Lise-Meitner Fellowship (Grant DOI 10.55776/M3300) 
 and 10.55776/P35673. TF gratefully acknowledges hospitality of Universit\'e Pierre et Marie Curie where parts of the project have been performed.

\newpage

\appendix
%\cleardoublepage
\widetext

The following is a series of appendices dedicated to technical details. In particular, Appendix \ref{Appendix_1} deals with Green's function on the cylinder both with and without external force, as well as the low-frequency expansion of propagators. In Appendix \ref{Appendix_2} we show how to express the variance of the displacement in terms of quantities accessible in the scattering formalism. Finally, Appendix \ref{sec_symmetry} provides numerous details involved in the calculations within the scattering formalism.

%%%%%%%%%%%%%%%%%%%%%%%%%%%%%%%%%%%%%%%%%%%%%%%%%%%%%%%%%%%%%%%%%%%%%
%%%%%%%%%%%%%%%%%%%%%%%%%%%%%%%%%%%%%%%%%%%%%%%%%%%%%%%%%%%%%%%%%%%%%
\section{Green's functions}
\label{Appendix_1}
In this appendix we will obtain the expression for the propagators in the domain with periodic boundary conditions. We will first consider the propagators in the absence of bias in Sec. \ref{sec_free_propagators}. The case of biased propagators will be examined in Sec.~\ref{sec_bias_propagators}.

%%%%%%%%%%%%%%%%%%%%%%%%%%%%%%%%%%%%%%%%%%%%%%%%%%%%%%%%%%%%%%%%%%%%%
%%%%%%%%%%%%%%%%%%%%%%%%%%%%%%%%%%%%%%%%%%%%%%%%%%%%%%%%%%%%%%%%%%%%%
\subsection{Green's functions on the cylinder}
\label{sec_free_propagators}
The explicit form of the propagator in the basis of lattice sites $|\textbf{r}\rangle$ can be obtained from the
conditional probability in the plane wave basis. The probability to observe the tracer at site $\textbf{r}$ at time $t$ knowing that it started its motion at position $\textbf{r}^{\prime}$ time $t=0$ is
\begin{align}
\langle \textbf{r} | \hat{U}_0(t) |\textbf{r}'\rangle = \frac{1}{N} \sum_{\textbf{k} \in \Lambda^*}
e^{\img\textbf{k}\cdot(\textbf{r}-\textbf{r}')} \exp[\epsilon(\textbf{k}) t] \, ,
\end{align}
where $\epsilon(\textbf{k}) = \epsilon_{x}(k_{x}) + \epsilon_{y}(k_{y})$, with $\epsilon_{x}(k) = \epsilon_{y}(k) = (-1/4)(1-\cos k)$ is the eigenvalue of the free Hamiltonian $\hat{H}_{0}$ associated to the bare dynamics, i.e., $\langle \textbf{k} \vert \hat{U}_{0}(t) \vert \textbf{k}^{\prime} \rangle = \exp[\epsilon(\textbf{k}) t] \delta(\textbf{k},\textbf{k}^{\prime})$, where $\delta(\textbf{k},\textbf{k}^{\prime})$ is the Kronecker delta. By virtue of translational invariance we can set $\textbf{r}^{\prime}= 0$ without loss of generality. In the limit of an infinitely long cylindrical lattice ($L_x\to\infty$) the sum over $k_x$ can be replaced by an integral $k_x \in [-\pi,\pi[$, thus we obtain the following representation
\begin{equation}
\label{prop1}
\begin{aligned}
\langle \textbf{r} | \hat{U}_0(t)  | \textbf{0}\rangle
 =& \frac{1}{L}\sum_{q=0}^{L-1} \exp\left( \frac{2\pi \img q y}{ L} + \epsilon_y( 2\pi q/L)t \right) \\
& \times \int_{-\pi}^\pi \frac{\diff k_x}{2\pi}
e^{\img k_x   x } \exp [\epsilon_x(k_x) t] \, .
\end{aligned}
\end{equation}
The conditional probability in the Laplace domain can be written as a resolvent of the hamiltonian $\hat{H}_0$ by virtue of the definition $\hat{G}_0(s) := \int_0^\infty \diff t\ e^{-s t} \hat{U}_0(t) = (s - \hat{H}_0)^{-1}$. In the basis of position kets, the above are known as lattice Green functions. For the case we are examining, we have
\begin{equation}
\begin{aligned}
\langle \textbf{r}|\hat{G}_0(s)|\textbf{0}\rangle & = \frac{1}{L} \sum_{q=0}^{L-1} \exp\left( \frac{2\pi \img q y}{ L} \right) \int_{-\pi}^\pi \frac{\diff k_x}{2\pi} \frac{e^{\img k_x x }}{s + 1 - [\cos(k_x) + \cos(2 \pi q/L)]/2 } \, .
\end{aligned}
\end{equation}
The integral can be performed relying on the residue theorem, a simple calculation gives
\begin{align}
\label{free_propagator}
\langle \textbf{r}| \hat{G}_0(s) |\textbf{0} \rangle &= \frac{2}{L}\sum_{q=0}^{L-1} \exp(  2\pi \img q y /L) \frac{ \left( 2s +2 - \cos(2\pi q/L) - \sqrt{ [2s + 2 - \cos(2\pi q/L)]^2 - 1} \right)^{|x|}}{ \sqrt{ [2s + 2 - \cos(2\pi q/L)]^2 - 1} } \, .
\end{align}

%%%%%%%%%%%%%%%%%%%%%%%%%%%%%%%%%%%%%%%%%%%%%%%%%%%%%%%%%%%%%%%%%%%%%
%%%%%%%%%%%%%%%%%%%%%%%%%%%%%%%%%%%%%%%%%%%%%%%%%%%%%%%%%%%%%%%%%%%%%
\subsection{Green's functions with biasing force}
\label{sec_bias_propagators}
Let us commence by considering the propagator in the time domain. The integral in (\ref{prop1}) can be identified as the integral representations of the modified Bessel function of the first kind of integer order $m$ \cite{gradshteyn_table_2014}
\begin{align}
I_m(z) = \int_{-\pi}^\pi \frac{\diff k}{2\pi}  e^{\img k m} \exp[ z \cos(k)] \, ;
\end{align}
hence, the conditional probability in the site basis (with no bias) is
\begin{equation}
\label{ }
\begin{aligned}
\langle \textbf{r}|\hat{U}_0(t)|\textbf{0}\rangle_{0} & = e^{-t/2}  I_{x}(t/2) \frac{1}{L}\sum_{q=0}^{L-1} \exp\left( \frac{2\pi \img q y}{ L} + \epsilon_y( 2\pi q/L)t \right) \, ;
\end{aligned}
\end{equation}
the subscript denotes the propagator without bias. Let us consider now the biased motion. The propagator can be found from (\ref{prop1}) and the integration over $k_{x}$ can be performed by using the following result
\begin{equation}
\label{ }
\int_{-\pi}^{\pi}\frac{\textrm{d}k}{2\pi} \, e^{\img k x + \alpha \cos k-\img\beta\sin k} = \left( \frac{\alpha+\beta}{\alpha-\beta} \right)^{\frac{x}{2}} I_{x}(\sqrt{\alpha^{2}-\beta^{2}}) \, ,
\end{equation}
where $I_{x}(\cdot)$ is a modified Bessel function of the first kind of order $x$. Taking $\alpha=(t/2)\cosh(F/2)$ and $\beta=(t/2)\sinh(F/2)$, the above formula yields
\begin{equation}
\label{prop_bias_free}
\begin{aligned}
\langle \textbf{r} | \hat{U}_0(t)  | \textbf{0}\rangle
& = \textrm{e}^{-t(\Gamma-1)} \textrm{e}^{Fx/2} \langle \textbf{r} | \hat{U}_0(t)  | \textbf{0}\rangle \big\vert_{F=0} \, ,
\end{aligned}
\end{equation}
where $\langle \textbf{r} | \hat{U}_0(t)  | \textbf{0}\rangle \big\vert_{F=0}$ is the propagator for the unbiased case ($F=0$) given by (\ref{prop1}). Passing to Laplace domain
\begin{equation}
\begin{aligned}
\langle \textbf{r}|\hat{G}_0(s)|\textbf{0}\rangle  =& \frac{\textrm{e}^{Fx/2}}{L} \sum_{q=0}^{L-1} \exp\left( \frac{2\pi \img q y}{ L} \right) \\
& \times \int_{0}^{\infty} \textrm{d}t \, \textrm{e}^{- r_{q}t/2} I_{x}(t/2) \, ,
\end{aligned}
\end{equation}
where $r_{q} = 2s+2\Gamma-2\epsilon_{y}(2\pi q/L)-1$, or equivalently, $r_{q} = 2s+2\Gamma-\cos(2\pi q/L)$; parenthetically, we observe that $r_{q}$ is a non-negative number, i.e., $r_{q}>2s+2\Gamma-1>2s$. The $t$-integral can be performed by using the identity
\begin{equation}
\int_{0}^{\infty} \textrm{d}z \, \textrm{e}^{-z \cosh\theta} I_{x}(z) = \frac{ \textrm{e}^{-x \theta }}{ \sinh\theta } \, .
\end{equation}
Summarizing, the biased propagator in Laplace domain is given by
\begin{equation}
\label{biased_propagator}
\begin{aligned}
\langle \textbf{r}|\hat{G}_0(s)|\textbf{0}\rangle & = \frac{2\textrm{e}^{Fx/2}}{L} \sum_{q=0}^{L-1} \exp \left( \frac{2\pi \img q y}{ L} \right) \frac{\Bigl[ r_{q} - (r_{q}^{2}-1)^{1/2} \Bigr]^{x}}{ (r_{q}^{2}-1)^{1/2} } \, .
\end{aligned}
\end{equation}
Note that when $F=0$ we have $r_{q}\vert_{F=0} = 2s+2-\cos(2\pi q/L)$ and (\ref{biased_propagator}) reduces to the free propagator given in (\ref{free_propagator}). The counterpart of (\ref{prop_bias_free}) in Laplace domain is implemented by shifting the Laplace parameter from $s$ to $s+\Gamma-1$, i.e.,
\begin{equation}
\label{ }
\langle \textbf{r}|\hat{G}_0(s)|\textbf{0}\rangle = \textrm{e}^{Fx/2} \left( \langle \textbf{r}|\hat{G}_0(s)|\textbf{0}\rangle \big\vert_{F=0, s \mapsto s+\Gamma-1} \right) \, .
\end{equation}
For convenience we single out the exponential factor in the above and we write the propagator as $\exp(Fx /2) g_{xy}(s)$.

%%%%%%%%%%%%%%%%%%%%%%%%%%%%%%%%%%%%%%%%%%%%%%%%%%%%%%%%%%%%%%%%%%%%%
%%%%%%%%%%%%%%%%%%%%%%%%%%%%%%%%%%%%%%%%%%%%%%%%%%%%%%%%%%%%%%%%%%%%%
\subsection{Low-frequency expansion}
\label{sec_low_freq_quasi_confined}
Here we show how to determine the low-$s$ behavior of propagators on the cylinder, focusing specifically on the quantity $\Delta_{L}(s)=4-\langle \textbf{0} | \hat{G}_0(s) |\textbf{0} \rangle+\langle 2\textbf{e}_{x} | \hat{G}_0(s) |\textbf{0} \rangle$. It is convenient to introduce $\psi_{q}>0$ defined as the positive root of
\begin{equation}
\label{ }
\cosh\psi_{q} = 2s+2-\cos(2\pi q/L) \, .
\end{equation}
Thanks to this definition the numerator in Eq.~(\ref{free_propagator}) simplifies considerably,
\begin{equation}
\label{ }
2s+2-\cos(2\pi q/L) - \sqrt{[2s+2-\cos(2\pi q/L)]^{2}-1} = \cosh\psi_{q} - \sinh\psi_{q} = \textrm{e}^{-\psi_{q}} \, .
\end{equation}
Analogously, the denominator is
\begin{equation}
\label{ }
\sqrt{ [2s + 2 - \cos(2\pi q/L)]^2 - 1} = \sinh\psi_{q} \, .
\end{equation}
The propagators we need are:
\begin{equation}
\label{ }
g_{00} = \frac{2}{L \sinh\psi_{0}} + \frac{2}{L} \sum_{q=1}^{L-1} \frac{1}{\sinh\psi_{q}} \, , \qquad g_{20} = \frac{2\textrm{e}^{-2\psi_{0}}}{L\sinh\psi_{0}} + \frac{2}{L} \sum_{q=1}^{L-1} \frac{\textrm{e}^{-2\psi_{q}}}{\sinh\psi_{q}} \, ,
\end{equation}
and therefore
\begin{eqnarray}
\label{13052025_1109}
\Delta_{L}(s) & = & 4 - \frac{4}{L} \textrm{e}^{-\psi_{0}} - \frac{4}{L} \sum_{q=1}^{L-1} \textrm{e}^{-\psi_{q}} \, .
\end{eqnarray}
The low-$s$ behavior of the first exponential in the above is
\begin{eqnarray} \nonumber
\textrm{e}^{-\psi_{0}} & = & \cosh\psi_{0} - \sinh\psi_{0} \\ \nonumber
& = & 1+2s - 2 \sqrt{s(s+1)} \\
& = & 1-2\sqrt{s} + O(s) \, .
\end{eqnarray}
Analogously, for the second exponential we have
\begin{equation}
\label{ }
\textrm{e}^{-\psi_{q}} = 2-c_{q} - \sqrt{3-4c_{q}+c_{q}^{2}} + O(s) \, , \qquad c_{q} : = \cos(2\pi q/L) \, .
\end{equation}
Recalling the identity $\sum_{q=1}^{L-1}c_{q}=-1$ it is immediate to find
\begin{equation}
\label{A14}
\Delta_{L}(s) = C_{L} + \frac{8}{L}\sqrt{s} + O(s) \, ,
\end{equation}
where $C_{L}$ is the confinement-dependent constant given by Eq.~(\ref{22092023_1208}); see Tab.~\ref{tab} for some values.
%Here some values
%\begin{equation}
%\begin{aligned}
%\label{}
%C_2 & = 4  \sqrt{2}- 4 \approx 1.65685 \\
%C_3 & = 4 \sqrt{7/3}-4\approx  2.1101 \\
%C_4 & = 2\sqrt{2}+ 2\sqrt{3} -4 \approx 2.29253 \\
%C_5 & = \frac{4}{5} \sqrt{35  + 2 \sqrt{205}}-4 \approx  2.38176 \\
%C_{10} & \approx  2.50474\\
%C_{100} & \approx  2.54606 \, .
%\end{aligned}
%\end{equation}
The limiting value $C_{\infty} := \lim_{L \rightarrow \infty} C_{L}$ is easily calculated
\begin{eqnarray} \nonumber
\label{29022024_1513}
C_{\infty} & = & -4 + \frac{2}{\pi} \int_{-\pi}^{\pi}dx \sqrt{3-4\cos x+\cos^{2} x} = \frac{8}{\pi} \, .
\end{eqnarray}
%\as{I corrected a typo: in Eq. (\ref{13062022_1330}) we have $C_{4}$ instead of $C_{3}$.}\\
We can actually push the expansion to order $O(s)$, a similar although rather lengthy calculation gives
\begin{equation}
\label{}
\Delta_{L}(s) = C_{L} + \frac{8}{L}\sqrt{s} + E_{L} s + O(s^{3/2}) \, ,
\end{equation}
with
\begin{equation}
\label{ }
E_{L} = -8 + \frac{ 8 }{ L } + \frac{ 8 }{ L } \sum_{q=1}^{L-1} \frac{ 2-\cos(2\pi q/L) }{ \sqrt{(2-\cos(2\pi q/L))^{2}-1} } \, .
\end{equation}
A careful analysis via Euler-Maclaurin is required in order to find how $E_{L}$ grows with $L$. Omitting the lengthy details, for large $L$ we find the asymptotic result
\begin{equation}
\label{ }
E_{L} = -4 + \frac{4}{\pi} + \frac{12}{\pi}\ln2 + \frac{8}{\pi}\left( \gamma-\frac{1}{2} + \ln\frac{L}{2\pi} \right) + O(L^{-1}) \, ,
\end{equation}
where $\gamma = 0.5772156...$ is Euler-Mascheroni constant. The above result implies that
\begin{equation}
\label{}
\Delta_{L}(s) = \frac{8}{\pi} + \left( - 4 + \frac{4}{\pi} + \frac{12}{\pi}\ln2 \right)s + \frac{8}{\pi}\left( \gamma-\frac{1}{2} + \ln\frac{L}{2\pi} \right)s + O(L^{-1}) \, .
\end{equation}
Notice that the first two terms in the above expression correctly recover the smooth part of the corresponding result obtained for the unbounded model [Eq.~(\ref{25092023_1254})]. In order to recover also the singular term $-(4/\pi)s\ln s$ one can perform a scaling limit such that $s \rightarrow 0$, $L \rightarrow \infty$ with the variable $\tilde{s} = sL^{2}$ fixed; for a detailed discussion on this aspect we refer the reader to Appendix \ref{scaling_diffusion_coefficient}.

%%%%%%%%%%%%%%%%%%%%%%%%%%%%%%%%%%%%%%%%%%%%%%%%%%%%%%%%%%%%%%%%%%%%%
%%%%%%%%%%%%%%%%%%%%%%%%%%%%%%%%%%%%%%%%%%%%%%%%%%%%%%%%%%%%%%%%%%%%%
\section{Variance of the displacement}
\label{Appendix_2}
The variance of the displacement is defined by $\textrm{Var}(t) = \langle \Delta x(t)^{2} \rangle -  \langle \Delta x(t) \rangle^{2}$. The first moment of the displacement can be decomposed into a contribution stemming from the empty lattice and a response one due to the obstacles. To first order in the density we have
\begin{equation}
\label{27022024_1512}
\langle \Delta x(t) \rangle = v_{0}t + n \int_{0}^{t}\textrm{d}t^{\prime} \, \Delta v(t^{\prime}) \, .
\end{equation}
An analogous decomposition is employed for the second moment, hence
\begin{equation}
\langle \Delta x(t)^{2} \rangle = 2D_{x}^{0}t + (v_{0}t)^{2} + n \Delta R_{x}(t) \, .
\end{equation}
To first order in $n$, the variance is
\begin{equation}
\textrm{Var}(t) = 2D_{x}^{0}t + n \biggl[ \Delta R_{x}(t) - 2 v_{0}t \int_{0}^{t}\textrm{d}t^{\prime} \, \Delta v(t^{\prime}) \biggr] \, .
\end{equation}
In the following we show how to express $\Delta v(t^{\prime})$ and $\Delta R_{x}(t)$ in terms of quantities that are accessible in the analytic framework formulated in Laplace domain. Passing to frequency domain and using the identity
\begin{equation}
\mathcal{L}\{ t \int_{0}^{t}\textrm{d}t^{\prime} \, g(t^{\prime}) \}(s) = \frac{1}{s^{2}}\left( 1-s\partial_{s} \right) \widehat{g}(s) \, ,
\end{equation}
the variance becomes
\begin{equation}
\widehat{\textrm{Var}}(s) = \frac{2D_{x}^{0}}{s^{2}} + n \biggl[ \widehat{\Delta R_{x}}(s) - 2 \frac{v_{0}}{s^{2}} \left( 1-s\partial_{s} \right) \widehat{\Delta v}(s) \biggr] \, .
\end{equation}
The rest of this section is devoted to express $\widehat{\Delta R_{x}}(s)$ and $\widehat{\Delta v}(s)$ in terms of the forward scattering amplitude in plane wave basis. To begin, we consider the velocity response, $\widehat{\Delta v}(s)$. Passing to frequency domain, the average displacement reads
\begin{eqnarray}\nonumber
\label{27022024_1510}
\mathcal{L}\{ \langle \Delta x(t) \rangle \}(s) & = & \img \partial_{k_{x}} [G(\textbf{k}) ]_{\rm av} \vert_{\textbf{k}=0} \\ \nonumber
& = & \img \partial_{k_{x}} [ G_{0} + n \left( G_{0} + Nt G_{0}^{2} \right) ] \vert_{\textbf{k}=0} \\ \nonumber
& = & \frac{v_{0}}{s^{2}} + n \widehat{\mathcal{W}}(s) \, ,
\end{eqnarray}
in the last line we used $\img \partial_{k_{x}}G_{0}\vert_{\textbf{k}=0}=v_{0}/s^{2}$ and we introduced 
\begin{equation}
\label{27022024_1526}
\widehat{\mathcal{W}}(s) := \img \partial_{k_{x}} \left( G_{0} + Nt G_{0}^{2} \right) \vert_{\textbf{k}=0} \, .
\end{equation}
Since $v(t) = (\textrm{d}/\textrm{d}t) \langle \Delta x(t) \rangle$, Eq.~(\ref{27022024_1510}) implies that
\begin{eqnarray} \nonumber
\label{27022024_1513}
\widehat{v}(s) & = & s \mathcal{L}\{ \langle \Delta x(t) \rangle \}(s) \\
& = & \frac{v_{0}}{s} + n s \widehat{\mathcal{W}}(s) \, .
\end{eqnarray}
By using the decomposition into bare contribution plus response [Eq.~(\ref{27022024_1512})], the average velocity in Laplace domain reads
\begin{eqnarray} \nonumber
\label{27022024_1514}
\widehat{v}(s) & = & \mathcal{L}\{ v(t) \}(s) \\ \nonumber
& = & \mathcal{L}\{ \partial_{t} \Delta \langle x (t)\rangle \}(s) \\ \nonumber
& = & \mathcal{L}\{ v_{0} + n \Delta v(t) \}(s) \\
& = & \frac{v_{0}}{s} + n \widehat{ \Delta v}(s) \, ,
\end{eqnarray}
by matching Eq.~(\ref{27022024_1513}) with Eq.~(\ref{27022024_1514}), we find
\begin{equation}
\label{ }
\widehat{ \Delta v}(s) = s \widehat{\mathcal{W}}(s) \, .
\end{equation}

We follow the very same guidelines for the second moment of the displacement. The scattering formalism allows us to identify the function $\widehat{\Delta R_{x}}(s)$ appearing in the obstacle-induced effects to the second moment of the displacement. From the disorder-averaged propagator we find the second moment
\begin{eqnarray} \nonumber
\mathcal{L}\{ \Delta \langle x(t) \rangle^{2} \}(s) & = & \img^{2}\partial_{k_{x}}^{2} [G_{\rm av}(\textbf{k})] \vert_{\textbf{k}=0} \\
& = & \frac{2D_{x}^{0}}{s^{2}} + \frac{2v_{0}^{2}}{s^{3}} + n \widehat{\Delta R_{x}}(s) \, ,
\end{eqnarray}
where
\begin{eqnarray}
\widehat{\Delta R_{x}}(s) & = &  \bigl[ - \partial_{k_{x}}^{2}G_{0} - 4 G_{0} (\partial_{k_{x}} G_{0}) (\partial_{k_{x}}Nt) - G_{0}^{2} (\partial_{k_{x}}^{2}Nt) \bigr] \vert_{\textbf{k}=0} \, .
\end{eqnarray}
By using the above expressions the variance in Laplace domain becomes
\begin{eqnarray} \nonumber
\widehat{\textrm{Var}}(s) & = & \frac{2D_{x}^{0}}{s^{2}} + n \biggl[ \widehat{\Delta R_{x}}(s) + 2 v_{0} \partial_{s}\widehat{\mathcal{W}}(s) \biggr] \, \\
& = & \frac{2D_{x}^{0}}{s^{2}} + n \widehat{Q}(s) \, ,
\end{eqnarray}
where in the last line we introduced $\widehat{Q}(s) := \widehat{\Delta R_{x}}(s) + 2 v_{0} \partial_{s}\widehat{\mathcal{W}}(s)$.

It turns out that both $\widehat{\Delta R_{x}}(s)$ and $\widehat{\mathcal{W}}(s)$ exhibit poles for $s=0$. The behavior at $s=0$ is captured by the Laurent expansion
\begin{equation}
\label{ }
\widehat{\Delta R_{x}}(s) = \frac{R_{3}}{s^{3}} + \frac{R_{2}}{s^{2}} + \frac{R_{1}}{s} + O(s^{0}) \, .
\end{equation}
An analogous representation holds true for the function appearing in the velocity response, i.e.,
\begin{equation}
\label{ }
\widehat{\mathcal{W}}(s) = \frac{W_{2}}{s^{2}} + \frac{W_{1}}{s} + O(s^{0}) \, .
\end{equation}
By combining the above results it follows that $\widehat{Q}(s)$ admits the Laurent expansion around $s=0$
\begin{equation}
\label{ }
\widehat{\mathcal{Q}}(s) = \frac{Q_{3}}{s^{3}} + \frac{Q_{2}}{s^{2}} + \frac{Q_{1}}{s} + O(s^{0}) \, ,
\end{equation}
where $Q_{3} = R_{3}-4v_{0}W_{2}$, $Q_{2} = R_{2}-2v_{0}W_{1}$. The third-order pole corresponds to a term $\propto t^{2}$ in time domain, which actually does not occur since $Q_{3}$ is identically zero by virtue of the identity $R_{3} = 4v_{0}W_{2}$. 

The asymptotic behavior of the variance at infinite times is fully captured by the low-frequency behavior, which is dominated by the term $Q_{2}$. This term actually does not vanish and leads to a linear growth in time and, as a result, this term is responsible for the obstacle-induced correction to the diffusion coefficient. The diffusion coefficient at infinite times is given by
\begin{equation}
\label{09122025_1039}
D_{x}(t \rightarrow \infty) = \lim_{s \rightarrow 0} s \widehat{D}(s) \, ,
\end{equation}
recalling that
\begin{equation}
\label{ }
\widehat{D}(s) = \frac{s}{2} \mathcal{L}\{ \textrm{Var}(t) \}(s) = \frac{s}{2} \widehat{\textrm{Var}}(s) \, ,
\end{equation}
we obtain
\begin{eqnarray}
D_{x}(t \rightarrow \infty) & = & D_{x}^{0} + \frac{n}{2} Q_{2} \\
& = & D_{x}^{0} + n \left( \frac{R_{2}}{2} - v_{0}W_{1} \right) \, .
\end{eqnarray}

In linear response we know that $\widehat{\mathcal{Q}}(s)$ exhibits simple poles for $s=0$ of the first and second order \cite{LF_2017}. For numerical purposes it is advantageous to employ a splitting of the function $\widehat{\mathcal{Q}}(s)$ into the sum of an analytic function and isolated poles. We therefore define the ``regularized'' $\widehat{\mathcal{Q}}$-function as
\begin{equation}
\label{ }
\widehat{\mathcal{Q}}^{(\rm reg)}(s) : =  \widehat{\mathcal{Q}}(s) - \frac{Q_{2}}{s^{2}} - \frac{Q_{1}}{s} \, ,
\end{equation}
hence; by construction, the function $\widehat{\mathcal{Q}}^{(\rm reg)}(s)$ vanishes at $s=0$. Passing to the time domain,
\begin{equation}
\label{ }
Q(t) = \widehat{\mathcal{Q}}^{(\rm reg)}(t) + Q_{2}t + Q_{1} \, ,
\end{equation}
where $\mathcal{Q}^{(\rm reg)}(t)$ formally denotes the inverse of the Laplace transform of $\widehat{\mathcal{Q}}(s)$; this step is implemented numerically by a suitable Filon-type algorithm. We can now assemble the result for the variance, which reads
\begin{eqnarray}
\textrm{Var}(t) & = & 2D_{x}^{0}t + n Q(t) \\
& = & 2 [ D_{x}^{0} + n Q_{2} ] t + n [ \mathcal{Q}^{(\rm reg)}(t) + Q_{1} ] \, ;
\end{eqnarray}
the first term yields the linear grow of the variance at large times. To conclude, we comment on the behavior of the term in square brackets, which is responsible for the non-trivial behavior of the complete time dependence. Since $\widehat{\mathcal{Q}}^{(\rm reg)}(s)$ is analytic in $s=0$ it follows that $\lim_{t \rightarrow \infty } \mathcal{Q}^{(\rm reg)}(t)=0$. The behavior for short times instead is captured by the limit $s \rightarrow \infty$ in frequency domain; for $t \rightarrow0$ we find that the square bracket vanishes.

%%%%%%%%%%%%%%%%%%%%%%%%%%%%%%%%%%%%%%%%%%%%%%%%%%%%%%%%%%%%%%%%%%%%%
%%%%%%%%%%%%%%%%%%%%%%%%%%%%%%%%%%%%%%%%%%%%%%%%%%%%%%%%%%%%%%%%%%%%%
\section{Scattering formalism in the symmetry-adapted basis}
\label{sec_symmetry}
%%%%%%%%%%%%%%%%%%%%%%%%%%%%%%%%%%%%%%%%%%%%%%%%%%%%%%%%%%%%%%%%%%%%%
%%%%%%%%%%%%%%%%%%%%%%%%%%%%%%%%%%%%%%%%%%%%%%%%%%%%%%%%%%%%%%%%%%%%%
\subsection{Orthogonal transformation}
In the absence of driving, $F=0$, all symmetries exhibited by the sing-obstacle problem are encoded in the dihedral group $D_{4}$ \cite{LF_2017}. These symmetries suggest the change of basis by means of the transformation induced by the orthogonal matrix
\begin{align}\label{eq:transM}
M = \begin{pmatrix}
    \frac{1}{\sqrt{2}} & 0 & \frac{1}{2} & \frac{1}{2\sqrt{5}} & \frac{1}{\sqrt{5}} \\
     0 &   \frac{1}{\sqrt{2}}  & -\frac{1}{2} & \frac{1}{2\sqrt{5}} & \frac{1}{\sqrt{5}} \\
    0 & 0 & 0 & -\frac{2}{\sqrt{5}} & \frac{1}{\sqrt{5}} \\
    0  & -\frac{1}{\sqrt{2}} & -\frac{1}{2} & \frac{1}{2\sqrt{5}} & \frac{1}{\sqrt{5}} \\
    -\frac{1}{\sqrt{2}} & 0 & \frac{1}{2} & \frac{1}{2\sqrt{5}} & \frac{1}{\sqrt{5}} 
\end{pmatrix} \, .
\end{align}
The new basis is provided by the column vectors that we denote as follows: $| p_y \rangle , | p_x \rangle ,  | d_{xy} \rangle , |s \rangle , | n \rangle$. Or, explicitly,
\begin{equation}
\begin{aligned}
\label{new_basis}
\sqrt{2} \vert p_{y} \rangle & = \vert \textbf{e}_{-2} \rangle - \vert \textbf{e}_{+2} \rangle \\
\sqrt{2} \vert p_{x} \rangle & = \vert \textbf{e}_{-1} \rangle - \vert \textbf{e}_{+1} \rangle \\
2 \vert d_{xy} \rangle & = \vert \textbf{e}_{-2} \rangle - \vert \textbf{e}_{-1} \rangle - \vert \textbf{e}_{+1} \rangle + \vert \textbf{e}_{+2} \rangle \\
2\sqrt{5} \vert s \rangle & = \vert \textbf{e}_{-2} \rangle + \vert \textbf{e}_{-1} \rangle - 4 \vert \textbf{e}_{0} \rangle + \vert \textbf{e}_{+1} \rangle + \vert \textbf{e}_{+2} \rangle \\
\sqrt{5} \vert n \rangle & = \vert \textbf{e}_{-2} \rangle + \vert \textbf{e}_{-1} \rangle + \vert \textbf{e}_{0} \rangle + \vert \textbf{e}_{+1} \rangle + \vert \textbf{e}_{+2} \rangle \, ,
\end{aligned}
\end{equation}
these vectors are linearly independent and normalized since $M M^{T}=1$, therefore they form a basis of $\mathbb{R}^{5}$. This notation aims at emphasizing the the reminiscent analogy of $\vert p_y \rangle$ and $\vert p_x \rangle$ as dipoles and $| d_{xy} \rangle$ as a vector with quadrupolar symmetry. Then, $|s \rangle$ corresponds to an $s$-wave and $| n \rangle$ is the neutral mode associated to conservation of probability since it is a left eigenvector of $\hat{v}$ with zero eigenvalue.

We can now translate the scattering problem due to a single obstacle in the new basis by writing $v^{\prime} = M^{T} v M$. In the absence of driving $v^{\prime}$ is the diagonal matrix $v^{\prime} = 4^{-1} \textrm{diag}(1,1,1,5,0)$; again, the vanishing of the last row reflects conservation of probability. In the presence of driving the matrix $v^{\prime}$ becomes
\begin{equation}
\label{ }
v^{\prime} = 
\left(\begin{array}{ccccc}
* & 0 & 0 & 0 & 0 \\
0 & * & * & *  & 2v_{0}/\sqrt{10} \\
0 & * & * & *  & 0 \\
0 & * & * & *  & 0 \\
0 & 0 & 0 & 0 & 0
\end{array}\right) \, ;
\end{equation}
where $*$ indicate a non-vanishing element. The free propagator becomes in the new basis, $G_{0}^{\prime} = M^{T} G_{0} M$, for $F=0$ is block-diagonal with the following structure
\begin{align}
G_{0}^{\prime} = \begin{pmatrix}
 * & 0 & 0 & 0 & 0  \\
 0 & * & \star & \star & \star  \\
 0 & \star & * & * & * \\
 0 & \star & * & * & *  \\
 0 & \star & * & * & *  
\end{pmatrix} \, ,
\end{align}
where $\star$ vanishes when $F=0$. We can already spot a difference between the quasi-confined model and the unbounded one. For the unbounded model at $F=0$ the free propagator $G_{0}^{\prime}$ is diagonal while for $F \neq 0$ the matrix $G_{0}^{\prime}$ is diagonal up to the non-vanishing elements $\langle s \vert G_{0}^{\prime} \vert n \rangle$ and $\langle n \vert G_{0}^{\prime} \vert s \rangle$. 

The scattering problem takes the very same form in the new basis (principle of covariance), in the new basis it reads
\begin{equation}
\label{ }
t^{\prime} = v^{\prime} + v^{\prime} G_{0}^{\prime} t^{\prime} \, .
\end{equation}
We will need the $S$-matrix, which is defined by $S = (1-vG_{0})^{-1}$. The orthogonal transformation acts on $S$ as follows
\begin{eqnarray} \nonumber
S & = & \left( 1 - vG_{0} \right)^{-1} \\ \nonumber
& = & \left( 1 - M v^{\prime} G_{0}^{\prime} M^{T} \right)^{-1} \\ \nonumber
& = & \Bigl[ M \left( 1 - v^{\prime} G_{0}^{\prime} \right) M^{T} \Bigr]^{-1} \\ \nonumber
& = & M \left( 1 - v^{\prime} G_{0}^{\prime} \right)^{-1} M^{T} \\
& = & M S^{\prime} M^{T} \, ,
\end{eqnarray}
therefore $S^{\prime} = M^{T} S M$ with $S^{\prime} = \left( 1 - v^{\prime} G_{0}^{\prime} \right)^{-1}$. We find
\begin{align}
(S^{\prime})^{-1} = (1- v' G_0') = 
\begin{pmatrix}
* & 0 & 0 & 0 & 0 \\
0 & * & \star & \star & \star \\
0 & \star & * & * & * \\
0 & \star & * & * & * \\
0 & 0 & 0 & 0 & 1 
\end{pmatrix} \, ,
\end{align}
and the transformed scattering matrix
\begin{align}
t' = (1- v' G_0')^{-1} v' = 
\begin{pmatrix}
* & 0 & 0 & 0 & 0 \\
0 & * & \star & \star & \star \\
0 & \star & * & * & \star \\
0 & \star & * & * & \star \\
0 & 0 & 0 & 0 & 0
\end{pmatrix} \, .
\end{align}

%%%%%%%%%%%%%%%%%%%%%%%%%%%%%%%%%%%%%%%%%%%%%%%%%%%%%%%%
\begin{comment}
\begin{equation}
\label{Sprime}
\renewcommand{\arraystretch}{0.70}
(S^{\prime})^{-1} = (1- v' G_0') =
\begin{tikzpicture}[baseline=(m.center)]
\matrix (m) [matrix of math nodes, left delimiter={(}, right delimiter={)}, row sep=-0.5mm, nodes={minimum width=0.0em, minimum height=0.0em}] {
* & 0 &  0 & 0 & 0 \\
0 & * &  \star & \star & \star \\
0 & \star &  * & * & * \\
0 & \star &  * & * & * \\
0 & 0 &  0 & 0 & 1 \\
      }; 
%      \node[highlight2, fit=(m-1-1.north west) (m-2-2.south east)] {};
\node[highlight1, fit=(m-3-3.north west) (m-4-4.south east)] {};
\end{tikzpicture} \, ,
\end{equation}
\end{comment}
%%%%%%%%%%%%%%%%%%%%%%%%%%%%%%%%%%%%%%%%%%%%%%%%%%%%%%%%

%%%%%%%%%%%%%%%%%%%%%%%%%%%%%%%%%%%%%%%%%%%%%%%%%%%%%%%%
\begin{comment}
\begin{equation}
\renewcommand{\arraystretch}{0.70}
t' = (1- v' G_0')^{-1} v' = 
\begin{tikzpicture}[baseline=(m.center)]
\matrix (m) [matrix of math nodes, left delimiter={(}, right delimiter={)}, row sep=-0.5mm, nodes={minimum width=0.0em, minimum height=0.0em}] {
* & 0 &  0 & 0 & 0 \\
0 & * &  \star & \star & \star \\
0 & \star &  t_{33}^{\prime} & t_{34}^{\prime} & \star \\
0 & \star &  t_{43}^{\prime} & t_{44}^{\prime} & \star \\
0 & 0 &  0 & 0 & 0 \\
      }; 
%      \node[highlight2, fit=(m-1-1.north west) (m-2-2.south east)] {};
\node[highlight1, fit=(m-3-3.north west) (m-4-4.south east)] {};
\end{tikzpicture} \, .
\end{equation}
\end{comment}
%%%%%%%%%%%%%%%%%%%%%%%%%%%%%%%%%%%%%%%%%%%%%%%%%%%%%%%%

The transformed basis and the basis $( \textbf{e}_{-2}, \dots, \textbf{e}_{2} )$ are related to each other via
\begin{equation}
\label{01032024_1343}
\vert \textbf{v}_{i} \rangle = M \cdot \vert \textbf{e}_{i} \rangle \, ,
\end{equation}
where $\vert \textbf{v}_{i} \rangle$ denotes the $i$-th vector of the new basis; e.g., $\vert \textbf{v}_{1} \rangle = \vert p_{y} \rangle$, $\vert \textbf{v}_{2} \rangle = \vert p_{x} \rangle$, $\vert \textbf{v}_{3} \rangle = \vert d_{xy} \rangle$. The inverse transformation is 
\begin{equation}
\label{05032024_1549}
\vert \textbf{e}_{i} \rangle = M^{T} \cdot \vert \textbf{v}_{i} \rangle \, .
\end{equation}
The entries of the transformed scattering-matrix $t^{\prime}$ can be expresses as certain matrix elements of the original $t$-matrix in the transformed basis. As an illustration, we consider the element $t_{22}^{\prime}$:
\begin{equation}
\begin{aligned}
\label{}
t_{22}^{\prime} & = \langle \textbf{e}_{2} \vert \hat{t}^{\prime} \vert \textbf{e}_{2} \rangle \\
& = \underbrace{\langle p_{x} \vert M}_{\langle \textbf{e}_{2} \vert } \underbrace{ M^{T} \hat{t} M }_{ \hat{t} } \underbrace{M^{T} \vert p_{x} \rangle}_{ \vert \textbf{e}_{2} \rangle } \\
& = \langle p_{x} \vert \hat{t} \vert p_{x} \rangle \, .
\end{aligned}
\end{equation}
To determine the matrix elements of the scattering $t$-matrix in the plane-wave basis, we only have to consider contributions from the distinguished subspace $\textbf{r} \in \{ \textbf{0} \} \cup \mathcal{N}$. Thus, we decompose the projection of the wave vector onto the distinguished subspace, $\hat{P} | \textbf{k}\rangle = N^{-1/2} \sum_{i} \exp( \img \textbf{k}\cdot \textbf{e}_{i} ) | \textbf{e}_{i} \rangle$. By using (\ref{05032024_1549}):
\begin{equation}
\label{01032024_1425}
\sqrt{N} \hat{P} \vert \textbf{k} \rangle =  \sum_{j} c_{j}(\textbf{k}) \vert \textbf{v}_{j} \rangle \, ,
\end{equation}
with
\begin{equation}
c_{j}(\textbf{k}) = \sum_{i} M_{ij} \textrm{e}^{\img \textbf{k} \cdot \textbf{e}_{i} } \, ,
\end{equation}
whose explicit expressions are:
\begin{equation}
\begin{aligned}
\label{}
c_{1}(\textbf{k}) & = - \sqrt{2} \img \sin(k_{y}) \\
c_{2}(\textbf{k}) & = - \sqrt{2} \img \sin(k_{x}) \\
c_{3}(\textbf{k}) & = \cos(k_{y}) - \cos(k_{x}) \\
c_{4}(\textbf{k}) & = \frac{1}{\sqrt{5}} \bigl[ \cos(k_{x}) + \cos(k_{y}) - 2 \bigr] \\
c_{5}(\textbf{k}) & = \frac{1}{\sqrt{5}} \bigl[ 2\cos(k_{x}) + 2\cos(k_{y}) + 1\bigr] \, .
\end{aligned}
\end{equation}
The projector then becomes
\begin{align} \nonumber
\sqrt{N} \hat{P} | \textbf{k} \rangle = & - \img \sqrt{2}\sin(k_y) | p_y \rangle - \img \sqrt{2}\sin(k_x) | p_x \rangle + [ \cos(k_y) - \cos(k_x) ] | d_{xy} \rangle \nonumber \\
&+ \frac{1}{\sqrt{5}} [ \cos(k_x) + \cos(k_y) -2 ] | s \rangle + \frac{1}{\sqrt{5}} [ 2 \cos(k_x) + 2 \cos(k_y) + 1 ]| n \rangle \, .
\end{align}
Let us consider now the forward-scattering matrix in the plane wave basis, $t(\textbf{k})= \langle \textbf{k} | \hat{t} | \textbf{k} \rangle$. From the above results
\begin{equation}
\begin{aligned}
\label{003}
N t(\textbf{k}) & = N \langle \textbf{k} \vert \hat{P} \hat{t} \hat{P} \vert \textbf{k} \rangle \\
& = N \langle \textbf{k} \vert \hat{P} (M \hat{t}^{\prime}M^{T}) \hat{P} \vert \textbf{k} \rangle \\
& = \left( \langle \textbf{k} \vert \hat{P} M \sqrt{N} \right) \hat{t}^{\prime} \left( \sqrt{N} M^{T} \hat{P} \vert \textbf{k} \rangle \right) \, ,
\end{aligned}
\end{equation}
where
\begin{equation}
\begin{aligned}
\label{proj}
\sqrt{N} M^{T} \hat{P} \vert \textbf{k} \rangle & = \sum_{i} c_{i}(\textbf{k}) \vert \textbf{e}_{i} \rangle \, ,
\end{aligned}
\end{equation}
and finally
\begin{equation}
\begin{aligned}
N t(\textbf{k}) & = \sum_{ij} c_{i}^{*}(\textbf{k})c_{j}(\textbf{k}) \langle \textbf{e}_{i} \vert \hat{t}^{\prime} \vert \textbf{e}_{j} \rangle \\
& = |c_{1}(\textbf{k})|^{2} \langle p_y  | \hat{t} | p_y \rangle + |c_{2}(\textbf{k})|^{2} \langle p_x  | \hat{t} | p_x \rangle + |c_{3}(\textbf{k})|^{2} \langle d_{xy} | \hat{t} | d_{xy} \rangle + |c_{4}(\textbf{k})|^{2} \langle  s | \hat{t} | s \rangle + 2 \textrm{Re}\bigl[ c_{3}(\textbf{k}) c_{4}^{*}(\textbf{k}) \bigr]  \langle s | \hat{t} | d_{xy} \rangle \, .
\end{aligned}
\end{equation}
The non-vanishing matrix elements are:
\begin{align}
\label{13062022_1332a}
\langle p_y  | \hat{t} | p_y \rangle & = \frac{1}{4 -g_{00} + g_{02} } \\
\label{13062022_1332b}
\langle p_x  | \hat{t} | p_x \rangle & = \frac{1}{4 -g_{00} + g_{20} } \\
\label{13062022_1332c}
\langle d_{xy} | \hat{t} | d_{xy} \rangle & = \frac{s}{4\Delta} \left( 1 - s g_{00} \right) \\
\label{13062022_1332d}
\langle  s | \hat{t} | s \rangle & = \frac{5}{4 \Delta} \left(s+2+g_{11}-g_{00} (s +1)^2\right) \\
\label{13062022_1332e}
\langle s | \hat{t} | d_{xy} \rangle & = \langle d_{xy} | t | s \rangle = \frac{s\sqrt{5}}{8\Delta} (  g_{01}-  g_{10} ) \, ,
\end{align}
the quantity
\begin{equation}
\begin{aligned}
\label{determinant}
\Delta \equiv \det\bigl[ (S^{\prime})^{-1}\vert_{sd} \bigr] = s [ g_{00}+  g_{11} - g_{01} - g_{10}   + s (g_{01} g_{10} - g_{00} g_{11} ) ] 
\end{aligned}
\end{equation}
is the determinant of the $2\times2$ matrix obtained by taking the subspace corresponding to the $d_{xy}$ and $s$ elements in the matrix $(S^{\prime})^{-1}$.

%%%%%%%%%%%%%%%%%%%%%%%%%%%%%%%%%%%%%%%%%%%%%%%%%%%%%%%%%%%%%%%%%%%%%%
%%%%%%%%%%%%%%%%%%%%%%%%%%%%%%%%%%%%%%%%%%%%%%%%%%%%%%%%%%%%%%%%%%%%%%
\subsection{Moments of the displacement}
\label{sec_moments}
The calculation of the moments is facilitated by using the following results. We can now list some useful properties of the projection operator that can be used for calculations that will follow. 
\begin{eqnarray}
\partial_{k_{x}} \sqrt{N} \hat{P} \vert \textbf{k} \rangle \big\vert_{ \bf{k}=\bf{0}} & = & - \sqrt{2} \img \vert p_{x} \rangle \\
\partial_{k_{y}} \sqrt{N} \hat{P} \vert \textbf{k} \rangle \big\vert_{ \bf{k}=\bf{0}} & = & - \sqrt{2} \img \vert p_{y} \rangle \\
\partial_{k_{x}}^{2} \sqrt{N} \hat{P} \vert \textbf{k} \rangle \big\vert_{ \bf{k}=\bf{0}} & = & \vert d_{xy} \rangle - \frac{1}{\sqrt{5}} \vert s \rangle  - \frac{2}{\sqrt{5}} \vert n \rangle \\
\partial_{k_{y}}^{2} \sqrt{N} \hat{P} \vert \textbf{k} \rangle \big\vert_{ \bf{k}=\bf{0}} & = & -\vert d_{xy} \rangle - \frac{1}{\sqrt{5}} \vert s \rangle  - \frac{2}{\sqrt{5}} \vert n \rangle \, .
\end{eqnarray}

Let us consider the velocity response. We consider
\begin{equation}
\label{ }
\frac{ \partial N t }{\partial k_{x} } \bigg\vert_{\textbf{k}=0} = \img \sqrt{10} \langle p_{x} \vert \hat{t} \vert n \rangle \, ,
\end{equation}
where $\langle p_{x} \vert \hat{t} \vert n \rangle = t_{25}^{\prime}$. This can be extracted from $t_{25}^{\prime} = (S^{\prime} v^{\prime})_{25}$, namely from the product of the second row of $S^{\prime}$ with the fifth column of $v^{\prime}$, the latter contains only one non-vanishing entry, $v_{25}^{\prime}=2v_{0}/\sqrt{10}$, thus $t_{25}^{\prime} = (2v_{0}/\sqrt{10}) S_{25}^{\prime}$. Therefore the function $\mathcal{V}_{L}(F;s)$ defined by $\img \partial_{k_{x}} N t \vert_{\textbf{k}=0} : = v_{0} \mathcal{V}_{L}(F;s)$ is given by $\mathcal{V}_{L}(F;s) = -2v_{0} S_{25}^{\prime}$. Although the full dependence on $F$ turns out to be rather long to be reported here, provided $s>0$, then for small $F$ we have
\begin{equation}
\label{ }
\mathcal{V}_{L}(F;s) = - \frac{8}{\Delta_{L}(s)} + O(F^{2}) \, ,
\end{equation}
where $\Delta_{L}(s)$ is defined in Eq.~\eqref{13052025_1109} and in Eq.~(\ref{11042024_1435}) it is evaluated for some values of $L$. For the sake of completeness, we mention that the small-$F$ expansion of $\mathcal{V}_{L}(F;s)$ at $s=0$ contains, beyond the $F$-independent term, a term of order $O(F)$. For the non confined model ($L \rightarrow \infty$) the result for $\mathcal{V}_{L}(F;0)$ is
\begin{eqnarray}
\mathcal{V}_{\infty}(F;0) & = & -2 + \frac{ (1-\Gamma) (2 \textsf{E}-\pi ) }{\textsf{E} - \left(1-\Gamma^{-1}\right) \textsf{K} -\pi/(2\Gamma)} + \frac{ \Gamma (2 \textsf{E}-\pi) }{ \textsf{E} - \left(1-\Gamma^{-2}\right) \textsf{K} } \, ,
\end{eqnarray}
the force $F$ enters through the complete elliptic integrals $\textsf{K} \equiv \textsf{K}[\Gamma^{-2}]$ and $\textsf{E} \equiv \textsf{E}[\Gamma^{-2}]$ with $\Gamma(F) = \cosh^{2}(F/4)$. For small forces the above admits the expansion
\begin{eqnarray}
\mathcal{V}_{\infty}(F;0) & = & - \pi + \frac{F^{2}}{32} [ (4-\pi ) \ln128 - 3 \pi ] + \frac{1}{16} (\pi -4) F^{2} \log|F| + \dots \, ,
\end{eqnarray}
this allows us to elaborate the corrections beyond FDT which, quite interestingly, are non-analytic in the driving force 
\begin{equation}
v(t \rightarrow \infty) = D_{x} F + \frac{n}{16} \left( \frac{\pi}{4} - 1 \right) F^{3}\ln|F| + O(|F|^{3}) \, ,
\end{equation}
with $D_{x}=[1-(\pi-1)n]/4$ the diffusion coefficient.

Finally, we consider the second moment of the displacement. We use
\begin{equation}
\begin{aligned}
\label{}
\partial_{k_{x}}^{2} N t(\textbf{k}) \big\vert_{\bf{k} = \bf{0}} & = 4 t_{22}^{\prime} + \sqrt{5} t_{35}^{\prime} - t_{45}^{\prime} \\
& = 4 \langle p_{x} \vert t^{\prime} \vert p_{x} \rangle + \sqrt{5} \langle d_{xy} \vert t^{\prime} \vert n \rangle - \langle s \vert t^{\prime} \vert n \rangle \, .
\end{aligned}
\end{equation}
We can obtain this result by observing that the only non-vanishing $\chi_{ij} = \partial_{k_{x}}^{2} c_{i}^{*} c_{j} \vert_{0}$ are: $\chi_{22} = 4$, $\chi_{35} = \chi_{53} = \sqrt{5}$, $\chi_{45} = \chi_{54} = -1$, and $\chi_{55} = -4$. The last however do not contribute since $t_{55}^{\prime}=0$, the same applies to $t_{53}^{\prime}=t_{54}^{\prime}=0$.

%%%%%%%%%%%%%%%%%%%%%%%%%%%%%%%%%%%%%%%%%%%%%%%%%%%%%%%%%%%%%%%%%%%%%%
%%%%%%%%%%%%%%%%%%%%%%%%%%%%%%%%%%%%%%%%%%%%%%%%%%%%%%%%%%%%%%%%%%%%%%
\subsection{Velocity response}
The complete time-dependent velocity response can be obtained by inverting the Laplace transform with the aid of Eq.~(\ref{05032024_1427}). In analogy with the analysis of the time-dependent diffusion coefficient, the procedure can be adopted by using the function $f(s)-f(t\rightarrow\infty)/s$ which, in time domain, tends to zero as $t \rightarrow \infty$. For the case at hand we consider $v(s)-v(t\rightarrow\infty)/s$. By using the expression for the terminal velocity [Eq.~(\ref{05032024_1425})] the approach of the velocity to the terminal value is given by
\begin{equation}
\label{05032024_1432}
v(t) - v(t \rightarrow \infty) = \frac{2n v_{0}}{\pi} \int_{0}^{\infty} \textrm{d}\omega \, \textrm{Re}\bigl[ g(\img \omega) \bigr] \cos(\omega t) \, ,
\end{equation}
where $g(s) : = [ \mathcal{V}_{L}(F;s) - \mathcal{V}_{L}(F;0) ]/s$. The analytical prediction for the normalized velocity response plotted in Fig.~\ref{fig2} is given by (\ref{05032024_1432}) upon normalizing to the initial value.

%%%%%%%%%%%%%%%%%%%%%%%%%%%%%%%%%%%%%%%%%%%%%%%%%%%%%%%%%%%%%%%%%%%%%%
%%%%%%%%%%%%%%%%%%%%%%%%%%%%%%%%%%%%%%%%%%%%%%%%%%%%%%%%%%%%%%%%%%%%%%
\subsection{Time-dependent diffusion coefficient}
\label{scaling_diffusion_coefficient}
The time-dependent diffusion coefficient in the frequency domain is given by
\begin{equation}
\label{03062024_0957}
\widehat{D}_{L}(s) = \frac{1}{4s} + n \biggl[ \frac{1}{s} \left( \frac{1}{4} - \frac{2}{\Delta_{L}(s)}\right) \biggr] \, ,
\end{equation}
where
\begin{equation}
\label{10062024_1019}
\Delta_{L}(s) = - 4 - 8 s - \frac{8}{L} \sqrt{s(1+s)} + \frac{4}{L} \sum_{q=0}^{L} \sqrt{ (2s+2-\cos(2\pi q /L))^{2}-1 } \, .
\end{equation}

In order to compute the inverse of the Laplace transform, we first single out the contribution yielding the terminal value of the time-dependent diffusion coefficient, which in Laplace domain takes the form $D_{x}^{\rm eq}/s$, where $D_{x}^{\rm eq} = \lim_{s \rightarrow 0} s \widehat{D}_{L}(s)$ is
\begin{equation}
\label{03062024_0958}
D_{x}^{\rm eq} = \frac{1}{4} + n \left( \frac{1}{4} - \frac{2}{\Delta_{L}(0)}\right) \, .
\end{equation}
We can now write Eq.~(\ref{03062024_0957}) in the form
\begin{equation}
\label{03062024_0959}
\widehat{D}_{L}(s) = \frac{D_{x}^{\rm eq}}{s} + \biggl[ \widehat{D}_{L}(s) - \frac{D_{x}^{\rm eq}}{s} \biggr] \, ,
\end{equation}
more explicitly, the above reads
\begin{equation}
\label{03062024_1000}
\widehat{D}_{L}(s) = \frac{D_{x}^{\rm eq}}{s} + n \biggl[ \frac{1}{s} \left( \frac{2}{\Delta_{L}(0)} - \frac{2}{\Delta_{L}(s)} \right) \biggr] \, .
\end{equation}
When converted into the time domain, the quantity in the square brackets becomes a function that vanishes for infinite times, therefore we can apply the technique outlined in the main text to find the time-dependent diffusion coefficient $D_{L}(t)$,
\begin{equation}
\label{03062024_1001}
D_{L}(t) = D_{x}^{\rm eq} + \frac{2n}{\pi} \int_{0}^{\infty} \textrm{d}\omega \, \textrm{Re}\left( F_{L}(\img \omega) \right) \cos(\omega t) \, ,
\end{equation}
where
\begin{equation}
\label{03062024_1002}
F_{L}(s) = \frac{1}{s} \left( \frac{2}{\Delta_{L}(0)} - \frac{2}{\Delta_{L}(s)} \right)
\end{equation}
encodes the obstacle-induced contribution to the time-dependent diffusion coefficient. We note that from the above relations we can find the VACF (recall that $\widehat{Z}_{L}(s) = s \widehat{D}_{L}(s)$) by taking a derivative with respect to time in Eq.~(\ref{03062024_1001}):
\begin{equation}
\label{03062024_1003}
Z_{L}(t) = - \frac{2n}{\pi} \int_{0}^{\infty} \textrm{d}\omega \, \omega \textrm{Re}\left( F_{L}(\img \omega) \right) \sin(\omega t) \, .
\end{equation}

\begin{figure}[htbp]
\centering
\includegraphics[width=110mm]{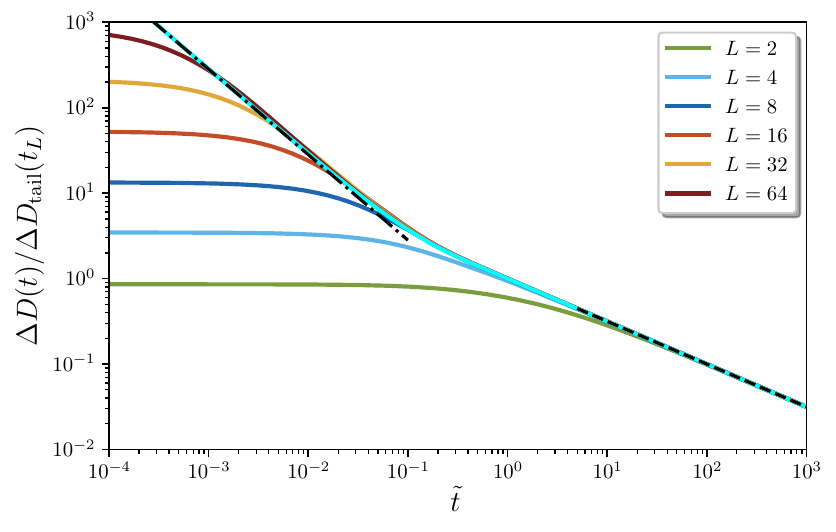}
\caption{The response to the time-dependent diffusion coefficient $\Delta D_{L}(t)$ rescaled to its asymptotic tail. The dashed black line indicates the exact long-time tail $\tilde{t}^{-1/2}$, the dot-dashed corresponds to the transient tail $c\tilde{t}^{-1}$, with $c=4/\pi^{3/2}$. The solid cyan is obtained from the scaling expression [Eq.~(\ref{03062024_1044}]) for a finite but large value of $L$; in this plot $L=2^{11}$.}
\label{fig_dscaling}
\end{figure}

Let us analyze the response contribution $\Delta D_{L}(t)$ defined by $D_{L}(t) = D_{x}^{\rm eq} + n \Delta D_{L}(t)$. From the low-frequency behavior of $\widehat{D}_{L}(s)$ we obtain the asymptotic form
\begin{equation}
\Delta D_{L}(t) \sim \frac{16}{\sqrt{\pi} L C_{L}^{2}} t^{-1/2} \equiv \Delta D_{\rm tail}(t) \, , \qquad t \rightarrow \infty \, .
\end{equation}
describing how $D_{L}(t)$ approaches the terminal value $D_{x}^{\rm eq}$. While a more profound analysis reveals a transient time tail of the form $D_{L}(t) \propto t^{-2}$ at shorter times and for large $L$. This feature suggests to seek for a scaling form for $D_{L}(t)$. By following the ideas already adopted for the VACF, we propose a scaling ansatz of the form
\begin{equation}
\label{03062024_1041}
\Delta D_{L}(t) = L^{-2} \mathcal{D}(t/t_{L}) \, ,
\end{equation}
for $t \rightarrow \infty$ and $L \rightarrow \infty$ with finite $t/t_{L}$, where $t_{L}=L^{2}$. It is simple to show that $L^{-2}$ in Eq.~(\ref{03062024_1041}) ensures the long-time behavior:
\begin{equation}
\label{03062024_1037}
\mathcal{D}(\tilde{t}) \propto
\begin{cases}
L^{-1} \tilde{t}^{-1/2} \, ,      & L < \infty, \\
\tilde{t}^{-1} \, ,      & L = \infty \, .
\end{cases}
\end{equation}
The plot of Fig.~\ref{fig_dscaling} provides the response to the time-dependent diffusion coefficient as a function of the rescaled time $t/L^{2}$. The response is measured in units of the asymptotic tail $\Delta D_{\rm tail}(t_{L})$. By inspection of this plot we conclude that for $L \rightarrow \infty$ a scaling function $\mathcal{D}(\tilde{t})$ characterized by two distinct power laws is actually emerging.
In the large-$L$ and large-$t$ limit, we expect that the scaling function to be given by
\begin{equation}
\label{03062024_1043}
\mathcal{D}(\tilde{t}) = \lim_{L \rightarrow \infty} \frac{2L^{2}}{\pi} \int_{0}^{\infty} \textrm{d}\omega \, \textrm{Re}\left( F_{L}(\img \omega) \right) \cos(\omega L^{2} \tilde{t}) \, ,
\end{equation}
a representation that follows from Eq.~(\ref{03062024_1001}) and the scaling ansatz [Eq.~(\ref{03062024_1041})]. This integral representation suggests to perform the change of variables $\Omega = \omega L^{2}$ in Eq.~(\ref{03062024_1043}), the latter yields a rather suggestive form for the scaling function $\mathcal{D}(\tilde{t})$:
\begin{equation}
\label{03062024_1044}
\mathcal{D}(\tilde{t}) = \lim_{L \rightarrow \infty} \frac{2}{\pi} \int_{0}^{\infty} \textrm{d}\Omega \, \textrm{Re}\left( F_{L}(\img \Omega/L^{2}) \right) \cos(\Omega \tilde{t}) \, ;
\end{equation}
however, this representation presents some subtleties due to the simplification of the overall $L^{2}$ outside the integral with the $L^{-2}$ stemming from $\textrm{d} \omega = L^{-2} \textrm{d} \Omega$. This is certainly correct provided $L$ is finite, while this operation must be handled with care if $L \rightarrow \infty$, in particular also the order of limit and integration is important.

%%%%%%%%%%%%
\section{The function $\mathcal{V}_{L}(F;s)$ for the quasi periodic two-lane model ($L=2$)}
\label{two_lane}
Here we provide the explicit expression of $\mathcal{V}_{2}(F;s)$. Since the result is rather long, we will write
\begin{equation}
\label{ }
\mathcal{V}_{2}(F;s) = \frac{ A(\cosh(F),\cosh(F/2),s) }{ B(\cosh(F),\cosh(F/2),s) } \, ,
\end{equation}
where $A(a,b,s)$ and $B(a,b,s)$ are the functions given by
\begin{align}
\label{}
\nonumber
A(a,b,s) & = \sqrt{2} b^2 \sqrt{(b+2 s-1) (b+2 s+3)} \sqrt{a+8 s (b+s)-1} + 2 \sqrt{2} b s \sqrt{(b+2 s-1) (b+2 s+3)} \sqrt{a+8 s (b+s)-1} \nonumber \\
& + 4 \sqrt{2} b \sqrt{(b+2 s-1) (b+2 s+3)} \sqrt{a+8 s (b+s)-1} - \sqrt{2} \sqrt{(b+2 s-1) (b+2 s+3)} \sqrt{a+8 s (b+s)-1} \nonumber \\
& -4 b^4-24 b^3 s-12 b^3-48 b^2 s^2-48 b^2 s-4 b^2 \sqrt{(b+2 s-1) (b+2 s+3)} \nonumber \\
& + 2 b^2 \sqrt{(b+2 s-1) (b+2 s+3)} \sqrt{(b+2 s+1) (b+2 s+3)} + 4 b^2-32 b s^3-48 b s^2+8 b s \nonumber \\
& -8 b s \sqrt{(b+2 s-1) (b+2 s+3)}+4 b s \sqrt{(b+2 s-1) (b+2 s+3)} \sqrt{(b+2 s+1) (b+2 s+3)} \nonumber \\
& +4 \sqrt{(b+2 s-1) (b+2 s+3)}-2 \sqrt{(b+2 s-1) (b+2 s+3)} \sqrt{(b+2 s+1) (b+2 s+3)}+12 b \, ,
\end{align}
and
\begin{align}
\label{}
\nonumber
B(a,b,s) & = 2 \sqrt{2} b^2 s \sqrt{a+8 s (b+s)-1} + 8 \sqrt{2} s^3 \sqrt{a+8 s (b+s)-1} + 8 \sqrt{2} b s^2 \sqrt{a+8 s (b+s)-1} \nonumber \\
&+ 16 \sqrt{2} s^2 \sqrt{a+8 s (b+s)-1} + 8 \sqrt{2} b s \sqrt{a+8 s (b+s)-1} + 6 \sqrt{2} s \sqrt{a+8 s (b+s)-1} \nonumber \\
& -6 b^3 s-4 b^3-36 b^2 s^2-42 b^2 s-2 b^2 s \sqrt{(b+2 s-1) (b+2 s+3)}+4 b^2 s \sqrt{(b+2 s+1) (b+2 s+3)} \nonumber \\
& +4 b^2 \sqrt{(b+2 s+1) (b+2 s+3)}-12 b^2-72 b s^3-8 s^3 \sqrt{(b+2 s-1) (b+2 s+3)} \nonumber \\
& +16 s^3 \sqrt{(b+2 s+1) (b+2 s+3)}-120 b s^2-8 b s^2 \sqrt{(b+2 s-1) (b+2 s+3)}-24 s^2 \sqrt{(b+2 s-1) (b+2 s+3)} \nonumber \\
&+16 b s^2 \sqrt{(b+2 s+1) (b+2 s+3)}+32 s^2 \sqrt{(b+2 s+1) (b+2 s+3)}-42 b s-12 b s \sqrt{(b+2 s-1) (b+2 s+3)} \nonumber \\
& -6 s \sqrt{(b+2 s-1) (b+2 s+3)}+24 b s \sqrt{(b+2 s+1) (b+2 s+3)}-4 s \sqrt{(b+2 s+1) (b+2 s+3)} \nonumber \\
& + 4 b \sqrt{(b+2 s+1) (b+2 s+3)}-8 \sqrt{(b+2 s+1) (b+2 s+3)}+4 b-48 s^4-104 s^3-36 s^2+26 s+12 \, .
\end{align}
The explicit expression of $\mathcal{V}_{L}(F;s)$ for arbitrary $L$ is more involved and can be obtained as illustrated in the supporting \texttt{Mathematica} script~\cite{LorentzGasNotebook}.
% \red{online repository available at [].}

\twocolumngrid

%\bibliographystyle{unsrt}
%\bibliographystyle{apsrev4-1-title}
%\bibliography{bibliography}

%%%%%%%%%%%%%%%%%%%%%%%%%%%%%%%%%%%%%%%%%%%%%%%%%%%%%%%%%%%%%%%%%%%%%
%%%%%%%%%%%%%%%%%%%%%%%%%%%%%%%%%%%%%%%%%%%%%%%%%%%%%%%%%%%%%%%%%%%%%

\end{document}